%% file: main.tex
\documentclass[aps,prd,twocolumn,superscriptaddress,groupedaddress,nofootinbib]{revtex4}
\usepackage{graphicx}  
\usepackage{dcolumn}   
\usepackage{bm}        
\usepackage{amssymb}   
\usepackage{amsmath}
\usepackage{tikz}
\usepackage{mathrsfs}
\usepackage{amsbsy}
\usepackage{mathtools}
\usepackage{appendix}
\usepackage{comment}
\DeclareMathAlphabet{\mathscrbf}{OMS}{mdugm}{b}{n}
\usepackage[svgnames]{xcolor} 
\usepackage{cancel}
\usepackage{aas_macros}
\usepackage{enumitem}

\usepackage[scr=boondox,  
            cal=esstix]   
           {mathalpha}
           
\allowdisplaybreaks

\newcommand{\wh}[1]{\widehat{#1}}

\usepackage[colorlinks=True, linkcolor=SteelBlue, citecolor=SteelBlue,urlcolor=SteelBlue]{hyperref}
\newcommand{\mbf}[1]{\mathbf{#1}}

\usepackage{orcidlink}

\newcommand{\param}{\ensuremath{\textsf{s}}}
\defcitealias{hegade26b}{I}

\begin{document}
\title{Nonlinear Perturbations of Relativistic Stars}
\author{K. J. Kwon\,\orcidlink{0000-0001-9802-362X}}
\email[]{jameskwon@ucsb.edu}
\affiliation{Department of Physics, University of California, Santa Barbara, CA 93106, USA}
\author{Abhishek Hegade K. R.}
\email[]{ah4278@princeton.edu}
\affiliation{Leinweber Forum For Theoretical Physics, Princeton University, Princeton, NJ 08544, USA}
\author{Tejaswi Venumadhav}
\email[]{teja@ucsb.edu}
\affiliation{Department of Physics, University of California, Santa Barbara, CA 93106, USA}
\affiliation{International Centre for Theoretical Sciences, Tata Institute of Fundamental Research, Bangalore 560089, India}
\author{Hang Yu\,\orcidlink{0000-0002-6011-6190}}
\email[]{hang.yu2@montana.edu}
\affiliation{eXtreme Gravity Institute, Department of Physics, Montana State University, Bozeman, MT 59717, USA}
\author{Nicol\'as Yunes}
\email[]{nyunes@illinois.edu}
\affiliation{Illinois Center for Advanced Studies of the Universe, Department of Physics, University of Illinois, Urbana-Champaign, Urbana, IL 61801, USA}
\begin{abstract}
We extend the general relativistic Lagrangian perturbation formalism of Friedman and Schutz~[Astrophys. J. \textbf{200}, 204 (1975)] beyond linear order, retaining full time dependence of the perturbations.
We provide general formulae to relate the Lagrangian and Eulerian perturbations to any general perturbative order.
As a proof of principle, we apply the framework to tidally perturbed relativistic stars in Regge--Wheeler gauge and reduce the second-order field and conservation equations to master equations with a nonlinear source.
Projecting the source onto the star's eigenmodes, we obtain a nonlinear equation of motion for the mode amplitudes and the coupling coefficients, which describe the interactions among the eigenmodes and with the external tidal field.
Approximating the tidal field and the coupling coefficients, we compute the static quadratic Love number from the quadrupolar fundamental mode alone and find that it reproduces the direct metric-matching value to a reasonable accuracy of $4\%$, up to a further correction arising from the Hamiltonian and momentum constraints of the second-order tidal field.
\end{abstract}
\maketitle
\section{Introduction\label{sec:intro}}
Neutron stars in coalescing binaries are tidally deformed due to the gravitational fields generated by their companion stars.
The deformation encodes information about the dense-matter equation of state~(see Ref.~\cite{chatziioannou25} for a review), while it also imprints itself in the observed gravitational-wave phases by altering the rate of the inspiral~\cite{flanagan08}.
Therefore, with an accurate model of the tidal phasing effect, gravitational-wave measurements of binary neutron-star mergers can constrain equation-of-state models of dense matter~\cite{gw170817_eos}.

A semi-analytical description of a neutron star's tidal response is obtained through a perturbative approach, in which a physical spacetime is described in terms of deviations from a background spacetime. The mapping between the physical and background spacetime is a diffeomorphism that can be parameterized using the \emph{strength} of the perturbation.
For a binary neutron star system, when the orbital frequency is well below the star's characteristic oscillation frequencies, the tidal response is driven on the orbital timescale.
In this regime, the strength of the externally forced perturbation is set by the instantaneous orbital separation, since the companion's tidal field falls off as separation increases.

When the perturbations are small and slowly varying, it suffices to describe a neutron star's tidal deformation as a time-independent linear perturbation.
In this case, the tidal deformation is simply proportional to the applied tidal field.
The proportionality constant is called the adiabatic tidal Love number~\cite{love09}, or equivalently, the tidal deformability~\cite{flanagan08, hinderer08,hinderer10}, which depends on the star's internal structure.
Formally, the tidal deformability modifies the gravitational-wave phases at 5th post-Newtonian~(PN) order, although its large magnitude makes the tidal effect potentially measurable.
It is through these coefficients that the dense-matter equation of state can be constrained through gravitational-wave observations~\cite{gw170817,gw170817_eos,gw170817_improve}.

As gravitational-wave emission drives the binary toward merger, the quasi-static description of the tidal response eventually becomes inadequate.
The magnitude of the tidal field increases and varies on timescales that can become comparable to the star's internal oscillation timescales. 

A realistic neutron star supports a rich spectrum of oscillation modes. 
Its bulk tidal response is dominated by the fundamental ($f$-) mode, with a characteristic frequency of order $2\,{\rm kHz}$, while additional modes associated with the star's microphysics occur at lower frequencies. 
These include gravity modes~\cite{reisenneger92} and interface modes~\cite{Counsell:2025hcv,pereira2025dynamicaltidesneutronstars}, whose characteristic frequencies are typically range from $100$ to $500\,\mathrm{Hz}$ for lowest-order modes.
Since these frequencies can lie within the range swept out by the orbital tidal forcing during the inspiral, such modes can be resonantly excited~\cite{lai94c}. 
Resonant excitation is intrinsically a dynamical phenomenon which cannot be captured by a static Love-number description.
Therefore, accurately modeling the hydrodynamic response of neutron stars during inspiral requires a framework that retains the full time dependence of the tidal interaction.

Moreover, as the binary shrinks and the tidal field grows stronger, nonlinear corrections to the stellar response can become increasingly important~\cite{yu23}. 
The leading nonlinear, time-independent tidal response is characterized by the quadratic Love number~\cite{poisson21,pitre25}, whose contribution to the gravitational-wave phase formally enters at 8th post-Newtonian order~\cite{pani25}. 
Despite its high PN order, this contribution is enhanced by a large numerical prefactor, much like the leading linear tidal correction. Ref.~\cite{pani25} showed that, within the static-tide approximation, the quadratic tidal contribution to the accumulated gravitational-wave phase can reach approximately $10\%$ of the linear tidal contribution near merger. 
Ref.~\cite{yu26} estimated that this second-order effect can be amplified by a factor of 2, solving a set of differential equations from Newtonian hydrodynamics with coefficients calibrated to match the low-frequency limit expected in general relativity.
A complete description of neutron-star tides in the late inspiral must therefore account for both the dynamical response of the star and nonlinear corrections to that response.

In Newtonian gravity, the theory of nonlinear and dynamical tides is well developed~\cite{dziembowski82,dziembowski85,wentzel87,dziembowski88,kumar89,kumar96,wu01,arras03,barker10,fuller12b,w12,w13,vzh,w16,yu23,kwon24,kwon25,yu26}. 
The stellar oscillation modes are obtained from a Hermitian pulsation operator~\cite{chandrasekhar64,lyndenbell1967} and, under suitable assumptions, form a complete basis for describing perturbations of the star. 
Each mode is associated with a well-defined overlap integral that determines how strongly it is driven by an external tidal field. 
The nonlinear perturbation equations can likewise be projected onto this mode basis, yielding coupling coefficients that quantify interactions among different oscillation modes. 
The tidal dynamics can therefore be reduced to a set of coupled mode-amplitude equations, with overlap integrals governing the external tidal forcing and coupling coefficients governing the nonlinear interactions among the modes.

In general relativity, the oscillations of relativistic stars have been studied extensively~\cite{thorne67,price69,thorne69a,thorne69b,campolattaro70,ipser73,detweiler73,detweiler75,lindblom83,detwiler85,lindblom97} (see Ref.~\cite{kokkotas99} for a review). 
Unlike the normal modes of a Newtonian star, gravitational radiation results in a set of quasi-normal modes; these quasi-normal modes arise from a non-Hermitian boundary-value problem and do not naturally come equipped with the complete orthogonal mode basis and well-defined tidal overlap integrals that underlie the Newtonian description. 

This difficulty has motivated a different approach at linear order, in which the tidal response is calculated directly without decomposing it into stellar modes.
Specifically, the exterior solutions describing the applied tidal field and the induced stellar response are obtained analytically and matched at the stellar surface to a numerical solution of the interior perturbation equations. The matching coefficients then determine the frequency-dependent tidal response. 
This strategy was first implemented within PN theory~\cite{HegadeKR:2024agt} and has subsequently been developed within effective-field-theory approaches~\cite{Andersson:2025iyd,Saketh:2026trm}. 
An alternative strategy is to compute the low-frequency response perturbatively and then resum the resulting expansion to recover the dominant $f$-mode contribution to the dynamical response~\cite{Pitre:2023xsr,Apostolidis:2026qsg}.

A complementary route at linear order is to recover the mode-based description familiar from Newtonian theory. 
In Refs.~\cite{hegade26a,hegade26b}, we developed a relativistic analogue of the overlap-integral formalism, allowing the tidal response to be related directly to the star's oscillation spectrum. 
Extending this description to nonlinear order requires coupling coefficients that characterize interactions among the modes.
More generally, however, nonlinear tidal dynamics need not be formulated in terms of modes at all: they require a systematic treatment of the coupled fluid and spacetime perturbations beyond first order.

In this paper, we develop a fully time-dependent nonlinear perturbation theory for relativistic perfect-fluid stars. 
Our construction is based on Lagrangian perturbation theory~\cite{friedman75,friedman78c,friedman13,bruni97}, which provides a natural description of the fluid degrees of freedom and allows effects such as stratification and time-dependent tidal forcing to be incorporated systematically. 
The resulting equations can be used directly to study the nonlinear tidal response, without introducing a mode decomposition. 
Alternatively, when combined with the relativistic mode formalism of Refs.~\cite{hegade26a,hegade26b}, they can be projected onto the stellar oscillation basis to derive relativistic mode-coupling coefficients and nonlinear mode-amplitude equations. 
Our treatment therefore provides both a general framework for nonlinear dynamical tides and a mode-based description of the underlying stellar dynamics.

As a first application and validation of the formalism, we specialize to the static nonlinear tidal problem and reproduce the quadratic tidal-response calculations of Refs.~\cite{poisson21,pitre25,pani25}. 
We further use this problem to test the mode-based description, finding agreement in the quadratic Love number with the direct calculation at the $\mathcal{O}(4\%)$ level. 
As a reference, at the linear order, the mismatch between the mode-sum expression of the Love number and direct matching was accurate to 3\%.
The fully time-dependent application of the formalism to nonlinear dynamical tides, together with a systematic assessment of the mode expansion across a broad range of neutron-star equations of state, will be presented in future work~\cite{nonlinear_dynamical_love_numbers,nonlinear-f-modes-gr}. 
The remainder of this section summarizes the framework and the main results of this paper.
We use geometric units throughout the paper and use the mostly plus metric signature.
\begin{figure*}
    \begin{minipage}[b]{\textwidth}
        \includegraphics[width=0.95\textwidth]{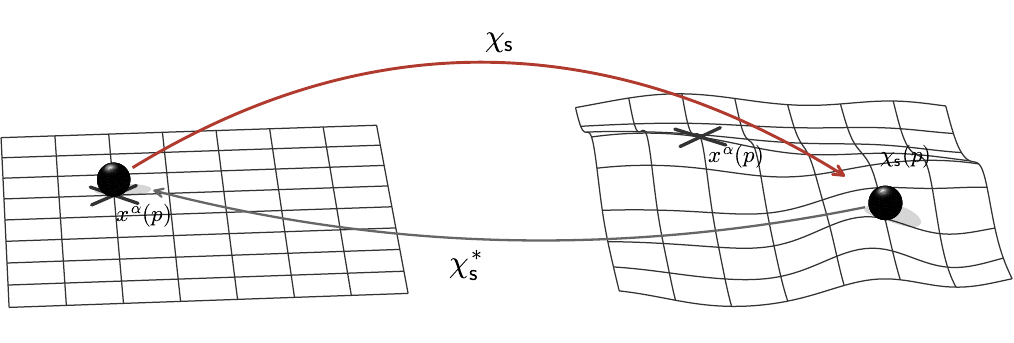}
        \caption{Visualization of the one-parameter family of diffeomorphism $\chi_{\ensuremath{\textsf{s}}}$ and pullback $\chi_{\ensuremath{\textsf{s}}}^\ast$.
        The fluid element (sphere) is located at $p$ (`x'-marker) in the background spacetime (left). 
        In the perturbed spacetime (right), the fluid element is displaced to  $\chi_{\ensuremath{\textsf{s}}}(p)$.}
        \label{fig:diffeo}
    \end{minipage}
\end{figure*}
\subsection*{Summary of Main Results\label{sec:exec_summary}}

\noindent \textbf{Lagrangian variations of the fluid variables.} 
The natural tool to describe perturbations of hydrodynamic variables is Lagrangian perturbation theory.
In Sec.~\ref{sec:formalism}, we explain the formalism needed for nonlinear Lagrangian perturbation theory at arbitrary order.
The main result of this section provides a general formula that relates Lagrangian and Eulerian perturbations at arbitrary order in perturbation theory [Eq.~\eqref{eq:relationship-between-lag-and-euler}]. 
We also display the explicit expressions for adiabatic Lagrangian variations in fluid thermodynamic variables through third order in perturbations.
These expressions are useful to compute the coupling coefficients and the nonlinear sources needed for higher-order perturbation theory.

\noindent \textbf{Nonlinear perturbations in Regge--Wheeler gauge and coupling coefficients.} 
In Sec.~\ref{sec:master-eqns}, we reduce the sourced field and conservation equations to master equations with a general source.
We provide the explicit expression for the second-order source in a supplementary \textsc{Mathematica} notebook, and the Eulerian four-velocity perturbation and the Lagrangian density perturbation in Eqs.~\eqref{eqn:eulerian_u_2nd_regge_wheeler} and \eqref{eqn:lag_density_regge_wheeler}.
In Sec.~\ref{sec:operator}, we review the operator formalism of Ref.~\cite{hegade26b}~(hereafter Paper~\citetalias{hegade26b}) to derive the sourced equation of motion of a stellar oscillation mode. 
In Sec.~\ref{sec:nonlinear-coupling}, we define the mode's triplet coupling coefficients in Eq.~\eqref{eqn:coupling_channel} and derive the second-order amplitude equation in Eq.~\eqref{eqn:second_amplitude}.

\noindent \textbf{Second-order static tidal response.}
In Sec.~\ref{sec:p2}, we apply our framework to study static, nonlinear, tidal perturbations of relativistic stars.
We express the static second-order response of the star to an external tidal field using oscillation modes and coupling coefficients in Eq.~\eqref{eqn:p2_full}.
We reduce it to the $f$-mode sum in Eq.~\eqref{eqn:p2_fmodes} and recover the Newtonian formula in the weak-field limit. 
We also present a numerical comparison with the static response calculated directly via metric-matching~\cite{pitre25} in Table~\ref{tab:p2_numbers}. 
Our mode sum differs by $\mathcal{O}(4\%)$ from the direct calculation, subject to the correction arising from the terms that impose Hamiltonian and momentum constraints on the second-order tidal field.
\section{Perturbation theory formalism\label{sec:formalism}}
In this section, we describe a general procedure to extend Lagrangian perturbation theory to arbitrarily high order, assuming the conservation of total baryon number and specific entropy.
In Sec.~\ref{sec:pre}, we set up the background spacetime and neutron star using the Tolman-Oppenheimer-Volkoff (TOV) equations;
Section~\ref{sec:lie} introduces the main technical concept of knight diffeomorphisms and provides the relationship between Lagrangian and Eulerian perturbations at arbitrary order in perturbation theory.
We then discuss how to integrate the density and energy conservation equations exactly to arbitrary order in perturbation theory in Sec.~\ref{sec:integrate-mass-euler}.
For an arbitrary perturbed quantity $Q$, its perturbation order will be denoted with a superscript as $Q^{(n)}$,
with the resummation convention
\begin{align}
    Q=\sum_{n}\frac{1}{n!}Q^{(n)}.
\end{align}
\subsection{Background Spacetime\label{sec:pre}}
Consider a spherically symmetric and static star in an equilibrium state. 
The geometry of the background manifold is described by the line element
\begin{align}
    ds^2=g_{\alpha\beta}dx^\alpha dx^\beta,
\end{align}
where the metric $g_{\alpha\beta}$ in Schwarzschild coordinate  $(t,r,\theta,\phi)$ is given by
\begin{align}
    g_{\alpha\beta}={\rm diag}(-e^{\nu},e^{\lambda}, r^2, r^2\sin^2\theta).\label{eqn:metric}
\end{align}
The stress-energy tensor of the perfect fluid within the star is
\begin{align}
    T^{\alpha\beta}=(\varepsilon + p)u^\alpha u^\beta + p g^{\alpha\beta}=\rho h u^\alpha u^\beta + p g^{\alpha\beta},\label{eqn:stress_energy}
\end{align}
where $\varepsilon=\varepsilon(r) $ is the energy density, $p=p(r) $ is the isotropic pressure, $h\equiv (\varepsilon + p)/\rho$ is the enthalpy per baryon, and $\rho$ is the baryon number density.
Also, $u^\alpha$ is the four-velocity of the fluid with the following normalization:
\begin{align}
    u^\alpha u_\alpha=-1,
\end{align}
which determines future-directed $u^t$ in the static equilibrium
\begin{align}
    u^t=e^{-\tfrac{\nu}{2}}.
\end{align}

The spacetime curvature sourced by the matter density satisfies Einstein's field equation
\begin{align}
    G^{\alpha\beta}=8\pi T^{\alpha\beta}.
\end{align}

The dynamics of the fluid within the star locally conserves energy
\begin{align}
    \nabla_\alpha T^{\alpha\beta}=0.
\end{align}
From the above, the components orthogonal and parallel to the fluid's four-velocity $u^\alpha$ respectively result in the relativistic Euler equation and the energy conservation equation. 
The motion of the fluid also conserves the baryon current
\begin{align}
    \nabla_\alpha(\rho u^\alpha)=0.
\end{align}

For the background spacetime, the equations of motion reduce to the Tolman-Oppenheimer-Volkoff (TOV) equations
\begin{subequations}
\begin{align}
     &\frac{dp}{dr}
     =
     -\rho h\frac{d\nu}{dr},\\
     &\frac{d\nu}{dr}
     =
     \frac{2}{r}\frac{m + 4\pi r^3 p}{r-2m},\\
     &\frac{dm}{dr}
     =
     4\pi  r^2 \varepsilon ,
\end{align}
\end{subequations}
where $m=m(r)$ is the enclosed mass.
Given a central energy density $\varepsilon_c$ and an equation of state $p(\varepsilon)$, we can integrate the TOV equations to obtain the background profiles, the radius $R$, the total mass $M\equiv m(R)$, and the compactness $C \equiv M/R$.
\subsection{Eulerian and Lagrangian Perturbations\label{sec:lie}}
We consider a one-parameter family of perfect-fluid spacetimes labeled by the parameter $\param$. For a tidally perturbed neutron star, $\param$ characterizes the strength of the tidal perturbation, with each value of $\param$ corresponding to an exact perfect-fluid solution. Fluid elements across this family of spacetimes are identified using a one-parameter family of diffeomorphisms $\chi_{\param}$, with
\begin{align}
    \chi_0=\mathrm{id}.
\end{align}
Heuristically, $\chi_{\param}$ maps a given fluid element in the background spacetime to the corresponding fluid element in the spacetime labeled by $\param$, as illustrated in Fig.~\ref{fig:diffeo}.

In coordinate form, the action of the pullback is particularly simple.
For a covariant tensor of rank $n$, we have
\begin{align}
    \left(
    \chi_{\param}^{*}\mathcal{T}_{\param}
    \right)_{\alpha_1\ldots\alpha_n}(x)
    &=
    \frac{\partial \chi_{\param}^{\mu_1}}{\partial x^{\alpha_1}}
    \cdots
    \frac{\partial \chi_{\param}^{\mu_n}}{\partial x^{\alpha_n}}
    \mathcal{T}_{\param,\mu_1\ldots\mu_n}
    \left(
    \chi_{\param}(x)
    \right)
    \,.
\end{align}
For a contravariant tensor, the inverse Jacobian appears.
For example, for a vector field $\mathcal{V}$,
\begin{align}
    \left(
    \chi_{\param}^{*}\mathcal{V}_{\param}
    \right)^{\alpha}(x)
    &=
    \frac{\partial x^{\alpha}}
    {\partial \chi_{\param}^{\mu}}
    \mathcal{V}_{\param}^{\mu}
    \left(
    \chi_{\param}(x)
    \right)
    \,.
\end{align}
Consequently, the pullback preserves tensor products and contractions. Therefore, algebraic tensor identities satisfied on the spacetime labeled by $\param$
remain valid after pulling the tensors back to the background spacetime.
In this section, for simplicity, we will suppress tensor indices.

Given a one-parameter family of diffeomorphisms, we can define two notions of change across the one-parameter family\footnote{An introduction to one-parameter family of diffeomorphisms and Lie derivatives can be found in Appendix C of Ref.~\cite{Wald-Textbook} and Appendix A of Ref.~\cite{friedman13}.} 
from the background ($\param = 0$). 
Consider a tensor field $\mathcal{T}_{\param}$ with the background tensor field denoted by
\begin{align}
    \mathcal{T} = \mathcal{T}_{\param=0}\,.
\end{align}
The Eulerian change of a tensor field $\mathcal{T}_{\param}$ is defined by
\begin{align}\label{eq:eulerian-change-def}
    \delta\mathcal{T}
    \equiv
    \mathcal{T}_\param-\mathcal{T}.
\end{align}
The Eulerian change compares the tensor fields at the same spacetime point as $\param$ is varied. From a fluid perspective, these points need not correspond to the same fluid element, since a perturbation generally displaces the fluid.
To compare the tensor evaluated on the same fluid element across the family of spacetimes, we introduce the Lagrangian change
\begin{align}\label{eq:Lagrangian-change-def}
    \Delta\mathcal{T}
    \equiv
    \chi_\param^*\mathcal{T}_\param-\mathcal{T},
\end{align}
where $\chi_\param^*\mathcal{T}_\param$ denotes the pullback of $\mathcal{T}_\param$ to the background spacetime. The pullback therefore allows tensors associated with corresponding fluid elements in different members of the family to be compared at the same point of the background spacetime.

We are now interested in characterizing the diffeomorphism $\chi_\param$. At linear order, we define the vector field
\begin{align}
    \xi^\mu
    =
    \left.
    \frac{d}{d\param}\chi_\param^\mu
    \right|_{\param=0},
\end{align}
which is tangent to the one-parameter family of maps at $\param=0$. Consequently,
\begin{align}
    \chi_\param^{\mu} (x)
    =
    x^{\mu}+\param \xi^{\mu}(x)+\mathcal{O}(\param^2).
\end{align}
The vector $\xi^\mu$ is called the Lagrangian displacement vector.
Using this equation in Eq.~\eqref{eq:Lagrangian-change-def}, we can show that
\begin{align}\label{eq:linear-perturbation-theory}
    \Delta\mathcal{T} = \delta \mathcal{T} + \param \mathcal{L}_{\xi} \mathcal{T} + \mathcal{O}(\param^2) \,,
\end{align}
where $\mathcal{L}$ denotes the Lie derivative.
The treatment of linear perturbation using the Lagrangian displacement vector is well studied in general relativity; see Ref.~\cite{friedman13} for a textbook treatment. 

There are several equivalent ways of representing the diffeomorphism $\chi_{\param}$ perturbatively beyond linear theory. 
Mathematically, this distinction arises because a generic one-parameter family of diffeomorphisms $\chi_{\param}$ cannot be generated for finite $\param$ by a single $\param$-independent vector field.
Such a construction would require $\chi_{\param}$ to satisfy the flow property
\begin{align}
    \chi_{\param + \textsf{t}}
    =
    \chi_{\param} \circ \chi_{\textsf{t}} \,,
\end{align}
which is not satisfied by a generic one-parameter family of diffeomorphisms; see Theorem 1 of Ref.~\cite{bruni97}.
Physically, this means that a generic finite nonlinear perturbation cannot simply be constructed by successively applying two smaller perturbations of the same form.
At linear order, perturbations superpose, and their effect is completely characterized by the tangent to the family at $\param = 0$.
Beyond leading order, however, successive perturbations interact with the perturbations that have already been applied.
Consequently, increasing the perturbation from $\param$ to $\param + \textsf{t}$ is not, in general, equivalent to first applying a perturbation of strength $\param$ and then applying the same map with strength $\textsf{t}$.

To extend the treatment of perturbations to second order and higher, we follow the approach of Ref.~\cite{bruni97}, which makes use of \emph{knight diffeomorphisms}~(see also Ref.~\cite{Friedman_2017}).
For different approaches to higher-order fluid perturbation theory, we refer readers to Refs.~\cite{friedman78a,Green_2014}.
In the framework of Ref.~\cite{bruni97}, the general diffeomorphism $\chi_{\param}$ is represented perturbatively as a composition of flows generated by a sequence of vector fields.
Each new vector field captures the additional information entering at the corresponding perturbative order.
Theorem 2 of Ref.~\cite{bruni97} provides a rigorous mathematical description of this statement. Here, we use this theorem and construct perturbative expansions of the diffeomorphism $\chi_{\param}$ and the Lagrangian change [Eq.~\eqref{eq:Lagrangian-change-def}] in terms of the Eulerian change [Eq.~\eqref{eq:eulerian-change-def}] and the Lie derivatives of the generating vector fields.

Let us begin by illustrating how one can obtain the second-order expression for the Lagrangian change.
Using Theorem 2 of Ref.~\cite{bruni97}, one can show that the diffeomorphism $\chi_{\param}$ can be generated using two successive flows of strength $\param$ and $\param^2/2$ generated by vector fields $\xi^{(1)}$ and $\xi^{(2)}$ respectively
\begin{align}
    \chi_{\param} = \chi^{(2)}_{\param^2/2} \circ \chi^{(1)}_{\param} + \mathcal{O}(\param^3) \,,
\end{align}
where $\chi^{(1)}_{\param}$ denotes the flow generated by vector field $\xi^{(1)}$ parameterized by $\param$.
Using Lemma~3 of Ref.~\cite{bruni97}, one can show that the diffeomorphism maps spacetime points to
\begin{align}
    \chi_{\param}^{\mu}(x)
    &=
    x^{\mu}
    +
    \param\, \xi^{(1)\mu}
    +
    \frac{\param^2}{2!}
    \Big[
    \xi^{(1)\nu} \partial_\nu\xi^{(1)\mu}
    +
    \xi^{(2)\mu}
    \Big]
    \nonumber\\
    &\hphantom{=}{}+
    \mathcal{O}\big(\param^3\big).
\end{align}
Similarly, using Lemma~2 of Ref.~\cite{bruni97} in Eq.~\eqref{eq:Lagrangian-change-def}, we obtain
\begin{align}
    \Delta \mathcal{T}
    =
    \delta \mathcal{T}
    +
    \param \mathcal{L}_{\xi^{(1)}}\mathcal{T}_{\param} + \frac{\param^2}{2!}\Big[\mathcal{L}_{\xi^{(1)}}^2 + \mathcal{L}_{\xi^{(2)}}\Big]\mathcal{T}_{\param}
    +
    \mathcal{O}(\param^3)
    \,.
\end{align}
Now, we use Eq.~\eqref{eq:eulerian-change-def}, and expand the Eulerian perturbation as
\begin{align}
    \delta \mathcal{T} \equiv 
    \sum_{n=1}^{\infty} 
    \frac{\param^n}{n!} \delta \mathcal{T}^{(n)} 
\end{align}
to $\mathcal{O}(\param^2)$, we obtain
\begin{align}
    \Delta \mathcal{T}
    &=
    \param \delta \mathcal{T}^{(1)}
    +
    \param \mathcal{L}_{\xi^{(1)}}\mathcal{T} 
    \nonumber\\
    &\quad + \frac{\param^2}{2!}
    \Big[
    \delta \mathcal{T}^{(2)}
    +
    \mathcal{L}_{\xi^{(1)}}^2 \mathcal{T} + \mathcal{L}_{\xi^{(2)}} \mathcal{T}
    +
    2 \mathcal{L}_{\xi^{(1)}} \delta T^{(1)}
    \Big]
    \nonumber\\
    &\quad +
    \mathcal{O}(\param^3).
\end{align}
To obtain the relationship between $\Delta \mathcal{T}$ and $\delta T^{(n)}$ we expand in the equation above in $\param$
\begin{align}\label{eq:DeltaT-expansion}
    \Delta \mathcal{T} \equiv 
    \sum_{n=1}^{\infty} 
    \frac{\param^n}{n!} \Delta \mathcal{T}^{(n)}
    \,,
\end{align}
and equate powers of $\param$ on the left and right-hand sides to get
\begin{subequations}
\begin{align}
    \label{eq:linear-reproduce-Lag-v-euler}
    &\Delta \mathcal{T}^{(1)} = 
    \delta \mathcal{T}^{(1)}
    +
    \mathcal{L}_{\xi^{(1)}}\mathcal{T} 
    \,,\\
    &\Delta \mathcal{T}^{(2)} = 
    \delta \mathcal{T}^{(2)}
    +
    \mathcal{L}_{\xi^{(1)}}^2 \mathcal{T} + \mathcal{L}_{\xi^{(2)}} \mathcal{T}
    +
    2 \mathcal{L}_{\xi^{(1)}} \delta \mathcal{T}^{(1)}
    \,.
\end{align}
\end{subequations}
Observe that Eq.~\eqref{eq:linear-reproduce-Lag-v-euler} is the same as Eq.~\eqref{eq:linear-perturbation-theory}.

The procedure outlined above can be generalized to arbitrary order in perturbation theory. 
From Lemma 2 of Ref.~\cite{bruni97},
\begin{align}\label{eq:lemma-2-eqn}
    \chi_{\param}^{*}\mathcal{T}_{\param}
    &=
    \sum_{\substack{i_1,\ldots,i_n \geq 0\\
    \sum_{j=1}^{n} i_j j \leq n}}
    \frac{
    \param^{\sum_{j=1}^{n} i_j j}
    }{
    \prod_{j=1}^{n} (j!)^{i_j}
    \prod_{j=1}^{n} i_j!
    }
    \mathcal{L}_{\xi^{(1)}}^{i_1}
    \cdots
    \mathcal{L}_{\xi^{(n)}}^{i_n}
    \mathcal{T}_{\param}
    \nonumber\\
    &\quad +
    \mathcal{O}(\param^{n+1})
    \,.
\end{align}
Using
\begin{align}
    \mathcal{T}_{\param}
    =
    \mathcal{T}
    +
    \sum_{k=1}^{n}
    \frac{\param^k}{k!}
    \delta \mathcal{T}^{(k)}
    +
    \mathcal{O}(\param^{n+1})
    \,,
\end{align}
and Eq.~\eqref{eq:DeltaT-expansion} in Eq.~\eqref{eq:lemma-2-eqn}, together with
\begin{align}
    \Delta \mathcal{T}
    =
    \chi_{\param}^{*}\mathcal{T}_{\param}
    -
    \mathcal{T}
    \,,
\end{align}
we get
\begin{widetext}
\begin{align}
    \sum_{k=1}^{n}
    \frac{\param^k}{k!}
    \Delta \mathcal{T}^{(k)}
    &=
    \sum_{\substack{i_1,\ldots,i_n \geq 0\\
    1 \leq \sum_{j=1}^{n} i_j j \leq n}}
    \frac{
    \param^{\sum_{j=1}^{n} i_j j}
    }{
    \prod_{j=1}^{n} (j!)^{i_j}
    \prod_{j=1}^{n} i_j!
    }
    \mathcal{L}_{\xi^{(1)}}^{i_1}
    \cdots
    \mathcal{L}_{\xi^{(n)}}^{i_n}
    \mathcal{T}
    \nonumber\\
    &\quad+
    \sum_{l=1}^{n}
    \sum_{\substack{i_1,\ldots,i_n \geq 0\\
    l+\sum_{j=1}^{n} i_j j \leq n}}
    \frac{
    \param^{l+\sum_{j=1}^{n} i_j j}
    }{
    l!
    \prod_{j=1}^{n} (j!)^{i_j}
    \prod_{j=1}^{n} i_j!
    }
    \mathcal{L}_{\xi^{(1)}}^{i_1}
    \cdots
    \mathcal{L}_{\xi^{(n)}}^{i_n}
    \delta \mathcal{T}^{(l)}
    +
    \mathcal{O}(\param^{n+1})
    \,.
\end{align}
Here, the term with $i_1=\cdots=i_n=0$ acting on $\mathcal{T}$ cancels the $-\mathcal{T}$ appearing in the definition of the Lagrangian change.
Equating the coefficients of $\param^n$ on both sides of the above equation, we get
\begin{align}
    \Delta \mathcal{T}^{(n)}
    &=
    \sum_{\substack{i_1,\ldots,i_n \geq 0\\
    \sum_{j=1}^{n} i_j j=n}}
    \frac{
    n!
    }{
    \prod_{j=1}^{n} (j!)^{i_j}
    \prod_{j=1}^{n} i_j!
    }
    \mathcal{L}_{\xi^{(1)}}^{i_1}
    \cdots
    \mathcal{L}_{\xi^{(n)}}^{i_n}
    \mathcal{T}
    \nonumber\\
    &\quad+
    \sum_{l=1}^{n}
    \sum_{\substack{i_1,\ldots,i_n \geq 0\\
    l+\sum_{j=1}^{n} i_j j=n}}
    \frac{
    n!
    }{
    l!
    \prod_{j=1}^{n} (j!)^{i_j}
    \prod_{j=1}^{n} i_j!
    }
    \mathcal{L}_{\xi^{(1)}}^{i_1}
    \cdots
    \mathcal{L}_{\xi^{(n)}}^{i_n}
    \delta \mathcal{T}^{(l)}
    \,.
\end{align}
The contribution with $l=n$ in the second term corresponds to $i_1=\cdots=i_n=0$ and gives $\delta \mathcal{T}^{(n)}$.
Therefore, the general relationship between the $n$th-order Lagrangian and Eulerian perturbations for any given tensor can be written as
\begin{align}\label{eq:relationship-between-lag-and-euler}
    \Delta \mathcal{T}^{(n)}
    &=
    \delta \mathcal{T}^{(n)}
    +
    \sum_{\substack{i_1,\ldots,i_n \geq 0\\
    \sum_{j=1}^{n} i_j j=n}}
    \frac{
    n!
    }{
    \prod_{j=1}^{n} (j!)^{i_j}
    \prod_{j=1}^{n} i_j!
    }
    \mathcal{L}_{\xi^{(1)}}^{i_1}
    \cdots
    \mathcal{L}_{\xi^{(n)}}^{i_n}
    \mathcal{T}
    \nonumber\\
    &\quad+
    \sum_{l=1}^{n-1}
    \sum_{\substack{i_1,\ldots,i_n \geq 0\\
    \sum_{j=1}^{n} i_j j=n-l}}
    \frac{
    n!
    }{
    l!
    \prod_{j=1}^{n} (j!)^{i_j}
    \prod_{j=1}^{n} i_j!
    }
    \mathcal{L}_{\xi^{(1)}}^{i_1}
    \cdots
    \mathcal{L}_{\xi^{(n)}}^{i_n}
    \delta \mathcal{T}^{(l)}
    \,.
\end{align}
\end{widetext}
The first few terms in the above equation are
\begin{align}
    \Delta \mathcal{T}^{(n)} = \delta \mathcal{T}^{(n)} 
    + 
    \mathcal{L}_{\xi^{(n)}} \mathcal{T}
    +
    n
    \mathcal{L}_{\xi^{(1)}}\mathcal{L}_{\xi^{(n-1)}} \mathcal{T}
    +
    \ldots
    \,.
\end{align}
\subsection{Perturbation Analysis}\label{sec:integrate-mass-euler}
The equations of motion for the Einstein-Euler system can be written as
\begin{subequations}
\begin{align}
    \boldsymbol{E}_{\param,\alpha\beta}
    &\equiv
    G_{\param,\alpha\beta}
    -
    8\pi T_{\param,\alpha\beta}
    =
    0
    \,,\\
    \boldsymbol{E}_{\param,\alpha}
    &\equiv
    \nabla_{\param,\beta}
    T_{\param,\alpha}{}^{\beta}
    =
    0
    \,.
\end{align}
\end{subequations}
Here, all quantities are evaluated on the member of the one-parameter family labeled by $\param$.
We now expand these equations perturbatively about the background solution at $\param=0$.

Schematically, at each order in perturbation theory, the equations take the schematic form
\begin{subequations}\label{eq:schematic-perturbation-equations}
\begin{align}
    &\boldsymbol{E}_{\alpha\beta}
    \left[
    h^{(n)}
    ,
    \xi^{(n)}
    \right]
    \nonumber\\
    &=
    \boldsymbol{S}_{\alpha\beta}^{(n)}
    \left[
    h^{(1)}
    ,
    \ldots
    ,
    h^{(n-1)}
    ,
    \xi^{(1)}
    ,
    \ldots
    ,
    \xi^{(n-1)}
    \right]
    \,,\\
    &\boldsymbol{E}_{\alpha}
    \left[
    h^{(n)}
    ,
    \xi^{(n)}
    \right]
    \nonumber\\
    &=
    \boldsymbol{S}_{\alpha}^{(n)}
    \left[
    h^{(1)}
    ,
    \ldots
    ,
    h^{(n-1)}
    ,
    \xi^{(1)}
    ,
    \ldots
    ,
    \xi^{(n-1)}
    \right]
    \,,
\end{align}
\end{subequations}
where 
\begin{align}
    h^{(n)}_{\alpha \beta} \equiv \delta g_{\alpha \beta}^{(n)}\label{eqn:eul_metric_def}
\end{align}
and the operators $\boldsymbol{E}_{\alpha\beta}$ and $\boldsymbol{E}_{\alpha}$ are the linearized Einstein and Euler operators evaluated on the background spacetime.
Consequently, the same linear differential operators act on the $n$th-order perturbations at every order in perturbation theory.
All nonlinearities are collected into the effective sources $\boldsymbol{S}_{\alpha\beta}^{(n)}$ and $\boldsymbol{S}_{\alpha}^{(n)}$, which depend only on perturbations of order lower than $n$.
At linear order these sources vanish, while at second and higher orders they describe the nonlinear coupling between lower-order perturbations.

Observe that the schematic equation in Eqs.~\eqref{eq:schematic-perturbation-equations} is derived by taking Eulerian perturbations. While metric perturbations can be naturally described in an Eulerian framework, we need to use the Lagrangian framework to relate Eulerian fluid perturbations to the gauge vector $\xi^{(n)}_{\alpha}$.
We can obtain these relations by solving the baryon number conservation equation, the entropy conservation, and adiabatic pressure variation conditions, evaluated in terms of the Lagrangian displacement and metric perturbation.

In this paper, we assume that our one-parameter family of diffeomorphisms conserves the total particle number and the specific entropy
\begin{align}\label{eq:number-and-specific-energy-conservation}
    \chi_{\param}^{*} \left( \rho_{\param} V_{\param} \right) = \rho V  \,,
    \quad
    \chi_{\param}^{*} \left( s_{\param} \right) = s
    \,,
\end{align}
where $V_{\mathrm{s}}$ is the proper spatial volume element in the local rest space of the fluid. 
These equations can be used to obtain the Lagrangian change in density, energy density, and pressure, as we demonstrate now.

Let us begin by understanding the consequence of the conservation of total particle number.
From Eq.~\eqref{eq:number-and-specific-energy-conservation} we have that
\begin{align}
    \Delta (\rho V ) = 
    \chi_{\param}^{*} ( \rho_{\param} V_{\param} ) - \rho V
    =
    0
    \,.
\end{align}
We can simplify this to
\begin{align*}
    (\Delta \rho + \rho) ( V + \Delta V ) - \rho V = 0
\end{align*}
which yields
\begin{align}
    \frac{\Delta \rho}{\rho}= -\frac{\Delta V}{V + \Delta V}
    \,.
    \label{eq:density-perturbation}
\end{align}
The change in the volume $\Delta V$ is given by
\begin{align}
    \frac{\Delta V}{V}=\frac{\Delta \sqrt{q}}{\sqrt{q}},\label{eqn:volume_pert}
\end{align}
where $q$ is the determinant of the induced metric on the constant-$t$ hypersurface in equilibrium. 
It also defines the projection tensor $q_{\alpha\beta}$ 
\begin{align}\label{eqn:proj_def}
    q_{\alpha\beta} \equiv u_{\alpha}u_{\beta} + g_{\alpha\beta}.
\end{align}
The right-hand side of Eq.~\eqref{eqn:volume_pert} can be evaluated in terms of $\Delta q_{\alpha \beta}$ following the method described in Appendix~\ref{appendix:lagrangian-variation}. 
The result for $\Delta \rho$ follows as\footnote{Note $\Delta q^{\alpha\beta}=\Delta (q^{\alpha\beta})\neq g^{\alpha\gamma}g^{\beta\delta}\Delta q_{\alpha\beta}$.}
\begin{align}
    &\frac{\Delta \rho}{\rho} = 
    -\frac{1}{2}q^{\alpha\beta}\Delta q_{\alpha\beta} 
    - \frac{1}{4}\bigg[\Delta q^{\alpha\beta} \Delta q_{\alpha\beta}  - \frac{1}{2} \big(q^{\alpha\beta}\Delta q_{\alpha\beta}\big)^2\bigg]
    \nonumber\\
    &
    +\frac{1}{8}\bigg[\Delta q^{\alpha\beta}\Delta q_{\alpha\beta}q^{\gamma\delta}\Delta q_{\gamma\delta}-\frac{1}{6}\big(q^{\alpha\beta}\Delta q_{\alpha\beta}\big)^3\bigg] 
     \nonumber \\
     &+\frac{1}{12}
     q^{\alpha\gamma}q_{\beta\epsilon}q^{\delta\zeta}\Delta q_{\alpha\beta}\Delta q_{\gamma\delta}\Delta q_{\epsilon\zeta} + \mathcal{O}(\Delta q^4) 
    \label{eqn:dn}
\end{align}

To derive the Lagrangian change in pressure, we recall the definition of the Lagrangian variation
\begin{align}
    \Delta p
    &=
    \chi_{\param}^{*} p_{\param}
    -
    p
    =
    p(
    \chi_{\param}^{*}\rho_{\param}
    ,
    \chi_{\param}^{*}s_{\param}
    )
    -
    p(\rho,s)
    \,.
\end{align}
Using Eq.~\eqref{eq:number-and-specific-energy-conservation} we obtain
\begin{align}
    \Delta p
    =
    p(
    \rho+\Delta\rho
    ,
    s
    )
    -
    p(\rho,s)
    \,.
\end{align}
On the right-hand side, expanding the first term in $\Delta\rho$ at fixed specific entropy gives
\begin{align}\label{eqn:lag_p_expr}
    \Delta p
    =
    \sum_{n=1}^{\infty}
    \frac{1}{n!}
    \bigg(
    \frac{\partial^n p}{\partial\rho^n}
    \bigg)_{s}
    (
    \Delta\rho
    )^n
    \,.
\end{align}
Using the definition of the adiabatic index 
\begin{align}\label{eq:adiabatic-index-def}
    \Gamma_1 \equiv \bigg(\frac{\partial \ln p}{\partial \ln \rho}\bigg)_s,
\end{align}
we can write the adiabatic Lagrangian variation in pressure as
\begin{align}
    &\frac{\Delta p}{p} 
    =\Gamma_1 \frac{\Delta \rho}{\rho} 
    + \frac{1}{2}\bigg[\Gamma_1(\Gamma_1-1)+ \bigg(\frac{\partial \Gamma_1}{\partial \ln \rho}\bigg)_s\bigg]\bigg(\frac{\Delta \rho}{\rho}\bigg)^2
    \nonumber\\
    &+\frac{1}{6}\bigg[\Gamma_1(\Gamma_1-1)(\Gamma_1-2)  + 3(\Gamma_1 -1)\bigg(\frac{\partial \Gamma_1}{\partial \ln \rho}\bigg)_s 
    \nonumber \\
    &+ \bigg(\frac{\partial^2 \Gamma_1}{\partial \ln \rho^2}\bigg)_s\bigg]\bigg(\frac{\Delta \rho}{\rho}\bigg)^3+
    \mathcal{O}(\Delta \rho ^4)
    \,.
\label{eq:lag-change-pressure}
\end{align}

We can obtain the change in energy density by following the same approach as above.
The Lagrangian change in the energy density is 
\begin{align}
    \Delta \varepsilon
    &=
    \chi_{\param}^{*} \varepsilon_{\param}
    -
    \varepsilon
    =
    \varepsilon(
    \chi_{\param}^{*}\rho_{\param}
    ,
    \chi_{\param}^{*}s_{\param}
    )
    -
    \varepsilon(\rho,s)
    \nonumber\\
    &=
    \varepsilon(
    \rho+\Delta\rho
    ,
    s
    )
    -
    \varepsilon(\rho,s)
    \nonumber\\
    &=
    \sum_{n=1}^{\infty}
    \frac{1}{n!}
    \bigg(
    \frac{\partial^n \varepsilon}{\partial\rho^n}
    \bigg)_{s}
    (
    \Delta\rho
    )^n
    \,.
\end{align}
The derivatives appearing in the equation above can be simplified using the first law of thermodynamics.
For a perfect fluid, the energy equation together with particle-number conservation implies
\begin{align}
    \left(
    \frac{\partial\varepsilon}{\partial\rho}
    \right)_s
    =
    \frac{\varepsilon+p}{\rho}
    \,.
\end{align}
Differentiating this relation with respect to $\rho$ at fixed $s$, and using the definition of the adiabatic index $\Gamma_1$~[Eq.~\eqref{eq:adiabatic-index-def}], we find
\begin{align}
    \bigg(
    \frac{\partial^2\varepsilon}{\partial\rho^2}
    \bigg)_s
    &=
    \frac{p\Gamma_1 }{\rho^2}
    \\
    \bigg(\frac{\partial^3\varepsilon}{\partial\rho^3}\bigg)_s
    &=
    \frac{p}{\rho^3}\bigg[
    \Gamma_1(\Gamma_1-2) 
    + \bigg(\frac{\partial\Gamma_1}{\partial \ln \rho}\bigg)_s\bigg].
\end{align}
Consequently, to cubic order, the Lagrangian change in the energy density is
\begin{align}
    \Delta\varepsilon
    &=
    \left(
    \varepsilon+p
    \right)
    \frac{\Delta\rho}{\rho}
    +
    \frac{1}{2}
    p\Gamma_1 
    \left(
    \frac{\Delta\rho}{\rho}
    \right)^2
    +\frac{1}{6}p
    \bigg[\Gamma_1(\Gamma_1-2)
    \nonumber\\
    &
    \quad
    +
    \bigg(\frac{\partial \Gamma_1}{\partial \ln \rho}\bigg)_s
    \bigg]
    \bigg(\frac{\Delta \rho}{\rho}\bigg)^3
    +
    \mathcal{O}(\Delta \rho^4)
    \,.
    \label{eq:lag-change-in-energy}
\end{align}

Equations~\eqref{eq:density-perturbation}, \eqref{eq:lag-change-pressure} and
\eqref{eq:lag-change-in-energy} completely determine the Lagrangian change in
density, pressure and energy density in terms of $\Delta q_{\alpha\beta}$.
Using Eq.~\eqref{eqn:proj_def},
the finite Lagrangian change in the spatial metric reads
\begin{align}
    \Delta q_{\alpha\beta}
    &=
    \Delta g_{\alpha\beta}
    +
    u_{\alpha}\Delta u_{\beta}
    +
    \Delta u_{\alpha}u_{\beta}
    +
    \Delta u_{\alpha}\Delta u_{\beta}
    \,,
\end{align}
where we can find $\Delta g_{\alpha\beta}$ from $\delta g_{\alpha\beta}$ using
Eq.~\eqref{eq:relationship-between-lag-and-euler}.

What remains is to determine the Lagrangian change in the fluid four-velocity.
We obtain it by following the argument leading up to Eq. (2.50) of Ref.~\cite{friedman13}.
Introduce an unnormalized tangent $\omega^\alpha$ to the fluid worldlines. Since $\chi_{\param}$ identifies the same fluid worldlines across the one-parameter family, we choose $\omega^\alpha$ such that
\begin{align}
    \chi_{\param}^{*}\omega_{\param}^{\alpha}
    =
    \omega^{\alpha}
    \,.
\end{align}
It follows that
\begin{align}
    \Delta \omega^{\alpha}
    =
    0
    \,.
\end{align}
The normalized four-velocity is
\begin{align}
    u^{\alpha}
    =
    \frac{\omega^{\alpha}}
    {\sqrt{-g_{\beta\gamma}\omega^{\beta}\omega^{\gamma}}}
    \,.
\end{align}
Using
\begin{align}
    \chi_{\param}^{*}g_{\param,\alpha\beta}
    =
    g_{\alpha\beta}
    +
    \Delta g_{\alpha\beta}
    \,,
\end{align}
we obtain
\begin{align}
    \chi_{\param}^{*}u_{\param}^{\alpha}
    &=
    \frac{\omega^{\alpha}}
    {\sqrt{
    -\left(
    g_{\beta\gamma}
    +
    \Delta g_{\beta\gamma}
    \right)
    \omega^{\beta}\omega^{\gamma}
    }}
    \nonumber\\
    &=
    u^{\alpha}
    \left(
    1
    -
    u^{\beta}u^{\gamma}
    \Delta g_{\beta\gamma}
    \right)^{-1/2}
    \,.
\end{align}
Therefore,
\begin{align}
    \Delta u^{\alpha}
    =
    u^{\alpha}
    \left[
    \left(
    1
    -
    u^{\beta}u^{\gamma}
    \Delta g_{\beta\gamma}
    \right)^{-1/2}
    -
    1
    \right]
    \,.
\end{align}
We expand the above using the generalized binomial series
\begin{align}
    \Delta u^{\alpha}=u^{\alpha}\sum_{n=1}^\infty\binom{-1/2}{n} (-1)^{n} \big(u^{\beta}u^{\gamma}\Delta g_{\beta\gamma}\big)^n.
\end{align}
At cubic order, this simplifies to
\begin{align}
    \Delta u^{\alpha}
    &=
    \frac{1}{2}
    u^{\alpha}u^{\beta}u^{\gamma}
    \Delta g_{\beta\gamma}
    +
    \frac{3}{8}
    u^{\alpha}
    \left(
    u^{\beta}u^{\gamma}
    \Delta g_{\beta\gamma}
    \right)^2
    \nonumber\\
    &
    \quad
    +
    \frac{5}{16}u^{\alpha}\big(u^{\beta}u^\gamma\Delta g_{\beta\gamma}\big)^3
    +
    \mathcal{O}(\Delta g^4)
    \,.\label{eqn:lag_u_3rd}
\end{align}
Finally, we can find $\Delta u_{\alpha}$ using
\begin{align}
    \Delta u_{\alpha}
    =
    g_{\alpha\beta}\Delta u^{\beta}
    +
    \Delta g_{\alpha\beta}u^{\beta}
    +
    \Delta g_{\alpha\beta}\Delta u^{\beta}
    \,.
\end{align}
\section{Second-Order perturbation theory\label{sec:second_order_perturbation_theory}}
In this section, we use the formalism developed in Sec.~\ref{sec:formalism} to study second-order polar perturbations in the Regge-Wheeler gauge. We first formulate the sourced master equations in Sec.~\ref{sec:master-eqns} and explicitly evaluate the perturbations entering these sources through second order.
We then recall some important results from Paper~\citetalias{hegade26b} in Sec.~\ref{sec:operator} which will be used to define formal coupling coefficients for the evolution of the amplitudes of the modes in Sec.~\ref{sec:nonlinear-coupling}.
\subsection{Sourced Master Equations in Regge-Wheeler Gauge}\label{sec:master-eqns}
Section~\ref{sec:integrate-mass-euler} provides necessary ingredients to evaluate the sourced Einstein-Euler equations in Eq.~\eqref{eq:schematic-perturbation-equations}.
After decomposing the perturbations into spherical harmonics, we will arrange the field equations into a closed system of ordinary differential equations for 4 perturbation variables.
We will refer to this reduced system as the master equations.

As discussed in Sec.~\ref{sec:formalism}, the differential operator acting on the perturbations has the same form at every perturbative order, while the nonlinearities are encoded in an effective source constructed from lower-order perturbations.
We therefore formulate the master equations for a general source $\bm{S}$.
The resulting master equations, written in terms of general source functions, are collected in Appendix~\ref{appendix:master_equation}.
In this section, we explicitly evaluate the perturbation variables and the corresponding source terms through second order in the Regge-Wheeler gauge.

We write the perturbed line element as  
\begin{align}
    ds^2
    &= 
    -
    e^{\nu}[1 -2 H(r) r^\ell  a(t)Y_{\ell m}]dt^2
    \nonumber \\
    &\quad -
    2i H_1(r) r^{\ell} Y_{\ell m} a(t)dt\,dr
    \nonumber \\
    &\quad +
    e^{\lambda}[1 + 2H_2(r) r^\ell  a(t)Y_{\ell m}]dr^2
    \nonumber \\
    &\quad +
    r^2[1-K(r) r^{\ell } a(t) Y_{\ell m} 
    ]
    d\Omega^2,
    \label{eqn:perturbed_line_element}
\end{align}
where $a(t)$ encodes the perturbation amplitude.
We also write the Lagrangian displacement vector $\xi^{\alpha}$ as
\begin{subequations}
\begin{align}
    \xi^t &=0,\\
    \xi^r &= r^{\ell-1}e^{-\lambda/2} W(r) a(t)Y_{\ell m},\\
    \xi^A &= -r^{\ell-2}V(r) a(t) \psi_{\ell m}^A.
\end{align}
\end{subequations}
The uppercase Latin index runs over the angular coordinates $\{\theta,\phi\}$ and $\psi^A_{\ell m}$ is the vector spherical harmonic of even parity
\begin{align}
    \psi^A_{\ell m} \equiv D^A Y_{\ell m}.
\end{align}
Here, $D$ is the covariant derivative associated with the unit 2-sphere metric 
$s^{AB}$.
For later convenience, we also define even-parity symmetric and tracefree~(STF) spherical harmonic tensor
\begin{align}
    \psi^{AB}_{\ell m} \equiv \bigg[ D^A D^{B}  + \frac{1}{2}\ell(\ell+1)s^{AB}\bigg]Y_{\ell m}.
\end{align}
Although we demonstrate our analysis for the even-parity perturbation, the odd-parity result follows using an odd-parity spherical harmonic basis.

Before proceeding, we note that the Regge-Wheeler gauge condition is well-defined for $\ell\geq 2$.
For even-parity perturbations, the choice of gauge vector eliminates $h_{tA}$, $h_{rA}$, and the traceless part of $h_{AB}$.
However, for radial perturbations, all of these components do not exist, whereas for dipole perturbations the traceless part of $h_{AB}$ is empty, which makes the gauge condition vacuous.
For $\ell=0$, we fix the gauge by choosing $\xi^\alpha$ such that
\begin{align}
    H_1=0, &&K=0, && V=0&&(\ell=0).\label{eqn:monopole_gauge_fixing}
\end{align}
This requires choosing $\xi^t$ in terms of $H_1$. However, a non-zero value of $\xi^t$ corresponds to the \emph{trivial displacement}~\cite{friedman78a}, which we set to zero again. 
For an $\ell=1$ perturbation, we set
\begin{align}
    K=0&&(\ell=1).\label{eqn:dipole_gauge_fixing}
\end{align} 

Using Eq.~\eqref{eqn:lag_u_3rd} and inverting Eq.~\eqref{eq:relationship-between-lag-and-euler}, the Eulerian perturbation of the four-velocity $\delta u^{\alpha}$ at the second order reads
\begin{widetext}
\begin{subequations}
\begin{align}
    \frac{\delta u^t}{Y_a} &= \frac{\widehat{H}_a}{e^{\nu/2 }}a_a 
    +
    \frac{1}{2 e^{\nu/2}}\sum_{bc}
    \bigg[
    3 T_{abc}   \wh{H}_{b}  \wh{H}_{c}    a_{b}  a_{c}
    -  
    \frac{i T_{abc}       }{ e^{\lambda/2  +  \nu    } r} 
    \big(
    \wh{H}_{1 c} \wh{W}_{b}\dot{a}_b a_{c}
    +
    \wh{H}_{1 b}    \wh{W}_{c} a_{b}  \dot{a}_c
    \big)
    +
    \frac{1}{e^\nu r^2}\big(F_{a}  \wh{V}_{b}   \wh{V}_{c}
    +
    T_{abc}  \wh{W}_{b}   \wh{W}_{c}\big)\dot{a}_b  \dot{a}_c\bigg],
    \\
    \frac{\delta u^r}{Y_a}&
    =
    \frac{ \wh{W}_{a} }{e^{(\lambda + \nu)/2 } r}\dot{a}_a
    + 
    \frac{1}{2 e^{(\lambda+\nu)/2}}\sum_{bc}
    \bigg\{
    \frac{T_{abc}}{  r} 
    \big(
    \wh{H}_{c}\wh{W}_{b}\dot{a}_b a_{c} 
    +
    \wh{H}_{b}    \wh{W}_{c} a_{b}  \dot{a}_c
    \big)
    +
    \frac{F_{a}}{r^3} \big(\wh{V}_{c}   \wh{W}_{b}   \dot{a}_b a_{c}  
    + 
    \wh{V}_{b}   \wh{W}_{c}   a_{b}\dot{a}_c\big)
    \nonumber\\
    &
    \quad
    + 
    \frac{1}{e^{\lambda/2}r^3}\big(F_{a}  \wh{V}_{b}   \wh{V}_{c}+T_{abc}  \wh{W}_{b}   \wh{W}_{c}\big)\big(\dot{a}_b  a_{c} + a_{b}  \dot{a}_c\big)
    -
    \frac{T_{abc}  }{e^{\lambda/2     } r^2} \big(\wh{W}_b'\wh{W}_{c}  \dot{a}_b a_{c}
    +
    \wh{W}_c'\wh{W}_{b} a_{b}\dot{a}_c
    \big)
    \bigg\},\\
    \delta u^A&=- \frac{ \wh{V}_a }{e^{\nu/2 } r^2}\dot{a}_a\psi^A_a+\frac{1}{2 e^{\nu/2}\Lambda_a^2}\psi_a^A\sum_{bc}
    \bigg\{
     -
     \frac{\wh{V}_{b}   \wh{V}_{c}     }{r^4} \big(G_{b}  \dot{a}_b a_{c}
     +
     G_{c} a_{b} \dot{a}_c
     \big)
     - \frac{1}{r^2}
     \big(
     F_{c}\wh{H}_{ c}    \wh{V}_{b}    \dot{a}_b a_{c}
     +
     F_{b}\wh{H}_{ b}    \wh{V}_{c}    a_{b} \dot{a}_c\big)
     \nonumber\\
     &\quad 
     +
     \frac{ 1}{ e^{\lambda/2 } r^4} \bigg[
     \big(
     F_{b}\wh{V}_{c}  \wh{W}_{b}   \dot{a}_b a_{c}   
     +
     F_{c}\wh{V}_{b}   \wh{W}_{c}   a_{b}   \dot{a}_c
     \big)
     -
     \big(
     F_{c}\wh{V}_{b}   \wh{W}_{c}    \dot{a}_b a_{c}    
     +
     F_{b}\wh{V}_{c}  \wh{W}_{b}  a_{b}   \dot{a}_c
     \big)
     +
     r
     \big(
     F_{c}   \wh{W}_{c}    \wh{V}_b'\dot{a}_ba_{c}
     +
     F_{b}    \wh{W}_{b}   \wh{V}_c'a_{b}\dot{a}_c
      \big)\bigg]
    \bigg\}.\label{eqn:euluA}
\end{align}
\label{eqn:eulerian_u_2nd_regge_wheeler}
\end{subequations}
where the lowercase Latin subscripts $a$, $b$, and $c$ implicitly indicate the angular quantum numbers $(\ell_a, m_a)$, $(\ell_b, m_b)$, and $(\ell_c, m_c)$, not mode indices.
For notational convenience, we have introduced
\begin{align}
     \wh{Q}_a &\equiv Q_a r^{\ell_a}, &&\Lambda_a^2 \equiv \ell_a(\ell_a + 1)
\end{align}
Also, $T_{abc}$, $F_a$, and $G_a$ are integrals involving three spherical harmonics, which we list in Appendix~\ref{appendix:spherical_harmonics_integral}.
We remind readers that the perturbation functions are subject to Eqs.~\eqref{eqn:monopole_gauge_fixing} and \eqref{eqn:dipole_gauge_fixing}.
Finally, Eq.~\eqref{eqn:euluA} holds for $\ell_a>0$ since $\Lambda_a^2=0$ for $\ell_a = 0$.
For a monopole perturbation, $\delta u^A$ instead vanishes.

Given the expression for the fluid's four-velocity, we are ready to evaluate the Lagrangian variation in fluid quantities via the relations presented in Sec.~\ref{sec:formalism}.
We have the following off-shell identity for the Lagrangian variation in the density
\begin{align}
    \frac{1}{Y_a}\frac{\Delta \rho}{ \rho}&=
    \bigg[
    -
    \wh{H}_{2 a}{}{} +  \wh{K}_{a}{}{}
    -  
    \frac{\Lambda_a^2{} \wh{V}_{a}{}{}}{r^2} 
    -
    \frac{1}{e^{\lambda/2}r^2}\big(\wh{W}_a + r\wh{W}_a'\big)
    \bigg] a_{a}
    +
    \sum_{bc} \bigg[e^{-\lambda/2}\wh{\mbf{P}}_b^T\,\mbf{M}_{bc}\,\wh{\mbf{P}}_c \,a_b a_c
    \nonumber\\
    &\quad 
    +
    \frac{i T_{abc}{}{}{}}{2 e^{\lambda/2 + \nu} r}  \big(\wh{H}_{1 b}{}{}  \wh{W}_{c} a_b \dot{a}_c+\wh{H}_{1 c}{}{} \wh{W}_{b}{}{}\dot{a}_b a_c\big)
    -
    \frac{1}{2e^{\nu}r^2}\big(F_{a}{} \wh{V}_{b}{}{} \wh{V}_{c}{}{} +  T_{abc}{}  \wh{W}_{b}{}{} \wh{W}_{c}{}{}\big) \dot{a}_b\dot{a}_c  \bigg],
    \label{eqn:lag_density_regge_wheeler}
\end{align}
where we have defined
\begin{align}
\wh{\mbf{P}}_a^T &
\equiv
\begin{pmatrix}
        \wh{H}_{1a} & \wh{H}_{2a} & \wh{K}_a & \wh{W}_a & \wh{V}_a &  \wh{H}_{2a}' & \wh{K}_a' & \wh{W}_a' & \wh{V}_a'
\end{pmatrix}
\end{align}
along with the coefficient matrix 
\begin{align}
\begingroup
\renewcommand*{\arraystretch}{1.5}
\mbf{M}_{bc}
=
\left(\begin{array}{ccccccccc}
\frac{T_{abc}}{2 e^{\lambda/2+\nu}} & 0 & 0 & 0 & 0 & 0 & 0 & 0 & 0
\\
0 & \tfrac{3}{2} e^{\tfrac{\lambda}{2}} T_{abc} & \ast & \ast & \ast & 0 & 0 & \ast & 0
\\
0 & -\tfrac{1}{2} e^{\tfrac{\lambda}{2}} T_{abc} & e^{\tfrac{\lambda}{2}} T_{abc} & \ast & \ast & 0 & 0 & \ast & 0
\\
0 & \frac{T_{abc}}{2 r^2} & -\frac{T_{abc}}{2 r^2} & \frac{T_{abc}\,(4 + r\lambda')}{2 e^{\lambda/2} r^4} & \ast & \ast & \ast & 0 & \ast
\\
0 & \frac{e^{\tfrac{\lambda}{2}} (F_{a} + \Lambda_b^2 T_{abc})}{2 r^2} & -\frac{e^{\tfrac{\lambda}{2}} (F_{a} + \Lambda_b^2 T_{abc})}{2 r^2} & \frac{F_{a} + 2 \Lambda_b^2 T_{abc}}{2 r^4} & \frac{2\bigl[F_{a} + e^{\lambda} (F_{a} + \Lambda_b^2 \Lambda_c^2 T_{abc} + U_{a})\bigr] + F_{a}\, r\lambda'}{4 e^{\lambda/2} r^4} & 0 & 0 & \ast & \ast
\\
0 & 0 & 0 & -\frac{T_{abc}}{2 r} & 0 & 0 & 0 & 0 & 0\\
0 & 0 & 0 & \frac{T_{abc}}{2 r} & 0 & 0 & 0 & 0 & 0\\
0 & \frac{T_{abc}}{2 r} & -\frac{T_{abc}}{2 r} & 0 & \frac{\Lambda_c^2 T_{abc}}{2 r^3} & 0 & 0 & \frac{T_{abc}}{e^{\lambda/2} r^2} & 0\\
0 & 0 & 0 & -\frac{F_{a}}{2 r^3} & -\frac{F_{a}}{2 e^{\lambda/2} r^3} & 0 & 0 & 0 & 0
\end{array}\right).
\endgroup
\end{align}
\end{widetext}
Here, $U_a$ is another integral involving three spherical harmonics (see Appendix~\ref{appendix:spherical_harmonics_integral}). 
The starred components can be inferred using 
\begin{align}
\mbf{M}_{bc}^T=\mbf{M}_{cb};
\end{align}
\emph{i.e.}, by taking $b\leftrightarrow c$.

Using Eqs.~\eqref{eq:relationship-between-lag-and-euler}, \eqref{eq:lag-change-pressure}, \eqref{eq:lag-change-in-energy}, and \eqref{eqn:lag_density_regge_wheeler}, we can derive the off-shell identities for pressure and energy density perturbations in Regge-Wheeler gauge.
Then, with the fluid perturbations evaluated, the perturbed Einstein equation can be arranged in the form of a system of four first-order ordinary differential equations~\cite{detwiler85} in terms of $H_1$, $K$, $W$, and a new variable $X$ defined via
\begin{align}
    \wh{X}&\equiv \frac{\ddot{a}(t)}{a(t)} e^{-\nu/2}\wh{V} + \frac{1}{r}\frac{1}{\rho h}\frac{dp}{dr}e^{(\nu-\lambda)/2}\wh{W} +e^{\nu/2}\wh{H}
    \label{eqn:x_definition}
\end{align}
Physically, $X$ is related to the adiabatic pressure perturbation at the linear order.

Notice that Eqs.~\eqref{eqn:perturbed_line_element}--\eqref{eqn:x_definition} are written in the time domain. We pass to the frequency domain using the Fourier transform
\begin{align}
    \tilde{a}(\omega)=\int e^{i\omega t}a(t)dt,
    \qquad
    a(t)=\frac{1}{2\pi}\int e^{-i\omega t}\tilde{a}(\omega) \,d\omega.
\end{align}
At linear order, each perturbation may be decomposed into harmonic components with time dependence
\begin{align}
a_a^{(1)}(t)\propto e^{-i\omega_a t}.
\end{align}
At second order, the perturbation is sourced by products of two linear-order perturbations. Consequently, a source proportional to
\begin{align}
a_b^{(1)}a_c^{(1)}
\propto e^{-i(\omega_b+\omega_c)t}
\end{align}
drives the second-order perturbation at the corresponding combination frequency,
\begin{align}
a_a^{(2)}(t)\propto e^{-i(\omega_b+\omega_c)t}.
\end{align}
Equivalently, multiplication in the time domain becomes a convolution in the frequency domain, so the second-order response at frequency \(\omega\) receives contributions from pairs satisfying \(\omega=\omega_b+\omega_c\).
The amplitude functions $a_a$, $a_b$, and $a_c$ in Eqs.~\eqref{eqn:perturbed_line_element}--\eqref{eqn:x_definition} represent resummed amplitudes $q^{(1)}+\tfrac{1}{2}q^{(2)}$ that contain both frequency-dependence.

In the frequency-domain, the second-order field equations reduce to a set of ordinary differential equations.
With $\mbf{Y}\equiv(H_1, K, W, X)^T$, the master equation takes a form of 
\begin{align}\label{eqn:master_eqn}
    \mbf{Y}'=\mbf{A} \cdot \mbf{Y} + \mbf{B}[\bm{S}].
\end{align}
The expression for the above equation with a general source $\bm{S}$ can be found in Appendix~\ref{appendix:master_equation}.

\subsection{Operator Formalism}\label{sec:operator}
Here, we describe the operator formulation of perturbation theory from Paper~\citetalias{hegade26b}. The goal is to organize the schematic equations from Eq.~\eqref{eq:schematic-perturbation-equations} into an integral operator form.
Since we are solving sourced, linearized perturbation equations on the TOV background at each order, we only need to understand the evolution of the gauge vector $\xi_{\mu} ^{(n)}$ and the metric perturbation $h_{\alpha \beta}^{(n)}$.
For simplicity, we drop the perturbation index $(n)$ and summarize the results of Paper~\citetalias{hegade26b}. 
Let us denote the linearized equations of motion schematically as
\begin{subequations}\label{eqn:field_eq_general_source}
\begin{align}
    &\boldsymbol{E}_{\alpha\beta}
    \left[
    \boldsymbol{y}
    \right]
    =
    \boldsymbol{S}_{\alpha\beta}
    \,,\\
    &\boldsymbol{E}_{\alpha}
    \left[
    \boldsymbol{y}
    \right]
    =
    \boldsymbol{S}_{\alpha}
    \,,
\end{align}
\end{subequations}
for some given source functions $\boldsymbol{S}_{\alpha \beta}$ and $\boldsymbol{S}_{\alpha}$. 
The symbol $\boldsymbol{y}$ denotes the collective fluid and metric perturbations
\begin{align}
    \boldsymbol{y}
    \equiv
    \left(
    h_{\alpha\beta}
    ,
    \xi^{\alpha}
    \right)
    \,.
\end{align}
We assume that the linearized vectors are Fourier decomposed with dependence $e^{-i \omega t}$ and 
consider the following parameterization for the metric perturbations
\begin{align}
    h_{\alpha \beta} &\equiv u_{\alpha} u_{\beta} \mathfrak{h} + i \omega u_{\alpha} \mathfrak{h}_{\beta} + i \omega u_{\beta} \mathfrak{h}_{\alpha} + \mathfrak{h}_{\alpha \beta}
    \,,
\end{align}
where we have suppressed the $e^{-i \omega t}$ dependence. Let us denote the Hamiltonian and momentum constraints as
\begin{subequations}\label{eq:linear-constraints}
\begin{align}
    \boldsymbol{H}[\boldsymbol{y}]
    &\equiv
    \boldsymbol{E}_{\mu\nu}[\boldsymbol{y}]
    u^{\mu}u^{\nu}
    \,,\\
    (-i\omega)\boldsymbol{P}_{\rho}[\boldsymbol{y}]
    &\equiv
    \boldsymbol{E}_{\mu\nu}[\boldsymbol{y}]
    u^{\mu}
    q^{\nu}{}_{\rho}
    \,,
\end{align}
\end{subequations}
To proceed further, we assume that the source functions do not impact the conservation of total baryon number and specific entropy.
We also demand that
\begin{align}\label{eq:constraints-on-source}
    \boldsymbol{S}_{\alpha \beta}u^{\alpha} u^{\beta} = 0 \,,
    \quad
    \boldsymbol{S}_{\alpha \beta}u^{\alpha} q^{\beta}{}_{\rho} = 0\,.
\end{align}
These conditions imply that the solution satisfies the Hamiltonian and momentum constraints.
We also fix the gauge freedom in the perturbations by assuming that the metric perturbations are in Regge-Wheeler gauge and the Lagrangian displacement vector satisfies
\begin{align}
    u_{\alpha} \xi^{\alpha} = 0\,.
\end{align}

Define the operator 
\begin{align}\label{eq:integration-v1}
    \boldsymbol{U}[\hat{\boldsymbol{y}}, \boldsymbol{y}] 
    &\equiv
    \int \sqrt{-g} d^3x
    \bigg[
    \hat{\xi}_{\beta} \boldsymbol{E}^{\beta}[\boldsymbol{y}] + \frac{\hat{h}_{\alpha \beta} \boldsymbol{E}^{\alpha \beta}[\boldsymbol{y}] }{16 \pi} 
    \nonumber\\
    &\quad -
    \frac{\hat{\omega} \omega}{8\pi} \boldsymbol{P}^{\rho}  \hat{\mathfrak{h}}_{\rho} +
    \frac{\omega^2}{8\pi}
    \hat{\boldsymbol{P}}^{\rho}
    \mathfrak{h}_{\rho}
    \bigg]
    \,,
    \nonumber\\
    &=
    \int \sqrt{-g} d^3x
    \bigg[
    \hat{\xi}_{\beta} \boldsymbol{S}^{\beta}+ \frac{\hat{h}_{\alpha \beta} \boldsymbol{S}^{\alpha \beta}}{16 \pi} 
    \bigg]
    \,.
\end{align}
By integrating by parts, the operator in Eq.~\eqref{eq:integration-v1} can be written in the form
\begin{align}\label{eq:operator-decomposition}
    \boldsymbol{U}
    \left[
    \hat{\boldsymbol{y}}
    ,
    \boldsymbol{y}
    \right]
    &=
    -
    \omega^2
    \boldsymbol{J}_0
    \left[
    \hat{\boldsymbol{y}}
    ,
    \boldsymbol{y}
    \right]
    +
    \boldsymbol{J}_1
    \left[
    \hat{\boldsymbol{y}}
    ,
    \boldsymbol{y}
    \right]
    +
    \left.
    \boldsymbol{\mathcal{R}}
    \left[
    \hat{\boldsymbol{y}}
    ,
    \boldsymbol{y}
    \right]
    \right|_{r=R}
    \,,
\end{align}
where $\boldsymbol{J}_0$ and $\boldsymbol{J}_1$ are frequency-independent bilinear operators and
$\boldsymbol{\mathcal{R}}$ is a boundary term evaluated at the surface of the star; see Paper~\citetalias{hegade26b} for explicit expressions.

The operator $\boldsymbol{U}$ can be used to define an inner product on the space of solutions that satisfy the near-zone boundary conditions defined in Paper~\citetalias{hegade26b}. 
Consider two solutions of the sourceless linearized equations,
$\boldsymbol{y}$ and $\hat{\boldsymbol{y}}$, with frequencies $\omega$ and $\hat{\omega}$.
Taking the antisymmetric combination of Eq.~\eqref{eq:operator-decomposition}, we define
\begin{align}
    \boldsymbol{E}[\hat{\boldsymbol{y}}, \boldsymbol{y}]
    &\equiv
    \boldsymbol{U}
    \left[
    \hat{\boldsymbol{y}}
    ,
    \boldsymbol{y}
    \right]
    -
    \boldsymbol{U}
    \left[
    \boldsymbol{y}
    ,
    \hat{\boldsymbol{y}}
    \right]
    \nonumber\\
    &=
    -
    \left(
    \omega^2
    -
    \hat{\omega}^2
    \right)
    \boldsymbol{I}_0
    \left[
    \hat{\boldsymbol{y}}
    ,
    \boldsymbol{y}
    \right]
    \,.
\end{align}
Here, $\boldsymbol{I}_0$ contains the bulk contribution from $\boldsymbol{J}_0$ together with the contribution obtained by symmetrizing the boundary term.
For the near-zone boundary conditions considered in Paper~\citetalias{hegade26b}, $\boldsymbol{I}_0$ is a symmetric bilinear operator,
\begin{align}
    \boldsymbol{I}_0
    \left[
    \hat{\boldsymbol{y}}
    ,
    \boldsymbol{y}
    \right]
    =
    \boldsymbol{I}_0
    \left[
    \boldsymbol{y}
    ,
    \hat{\boldsymbol{y}}
    \right]
    \,.
\end{align}
Since both perturbations satisfy the sourceless equations of motion,
\begin{align}
    \boldsymbol{E}[\hat{\boldsymbol{y}}, \boldsymbol{y}]
    =
    0
    \,,
\end{align}
and hence
\begin{align}
    \left(
    \omega^2
    -
    \hat{\omega}^2
    \right)
    \boldsymbol{I}_0
    \left[
    \hat{\boldsymbol{y}}
    ,
    \boldsymbol{y}
    \right]
    =
    0
    \,.
\end{align}
The stellar eigenmodes are therefore orthogonal with respect to $\boldsymbol{I}_0$.
Denoting the eigenmodes by $\boldsymbol{y}_s$ with frequencies $\omega_s$, we have
\begin{align}\label{eq:mode-orthogonality}
    \boldsymbol{I}_{0} \left[\bm{y}_{p}, \bm{y}_{s}\right] = \frac{R M^2}{(R \omega_{s})^2} \delta_{sp} \mathscr{N}_{s}
    \,,
\end{align}
where $\mathscr{N}_s$ denotes the normalization of mode $s$.

We now return to the sourced problem.
Since the solution satisfies the sourceless Hamiltonian and momentum constraints, we assume that it can be expanded in the eigenmodes of the star as
\begin{align}
    \boldsymbol{y}
    =
    \sum_s
    a_s(\omega)
    \boldsymbol{y}_s e^{-i \omega t}
    \,.
\end{align}
Substituting this expansion into Eq.~\eqref{eq:integration-v1} and choosing
$\hat{\boldsymbol{y}}=\boldsymbol{y}_p$, we obtain
\begin{align}
    &\sum_s
    a_s(\omega)
    \left(
    \omega_p^2
    -
    \omega^2
    \right)
    \boldsymbol{I}_0
    \left[
    \boldsymbol{y}_p
    ,
    \boldsymbol{y}_s
    \right]
    \nonumber\\
    &=
    \int
    \sqrt{-g}
    \left[
    \xi_{p,\beta}
    \boldsymbol{S}^{\beta}
    +
    \frac{
    h_{p,\alpha\beta}
    \boldsymbol{S}^{\alpha\beta}
    }{16\pi}
    \right]
    d^3x
    \,.
\end{align}
Using the orthogonality relation in Eq.~\eqref{eq:mode-orthogonality}, the amplitude of mode $p$ is therefore determined by
\begin{align}
    a_p (\omega)
    &=
    \frac{1}{\mathscr{N}_p \left( 1 - \omega^2/\omega_p^2 \right)}
    \frac{R}{M^2}
    \int
    \nonumber \\
    &\quad \times \sqrt{-g}
    \left[
    \xi_{p,\beta}
    \boldsymbol{S}^{\beta}
    +
    \frac{
    h_{p,\alpha\beta}
    \boldsymbol{S}^{\alpha\beta}
    }{16\pi}
    \right]
    d^3x
    \,,
\end{align}
for a general source $\bm{S}$ that satisfies Eq.~\eqref{eq:constraints-on-source}.
Thus, the relativistic mode amplitude takes the familiar harmonic-oscillator form, with the forcing determined by the projection of both the fluid and gravitational source terms onto the stellar eigenmode.
However, a few important differences are worth noting here. In our approach, the operator $\boldsymbol{I}_{0}$ is not positive definite. Therefore, the mode expansion is only a formal expansion, and it does not converge exactly to the true solution as shown in Paper~\citetalias{hegade26b}. 
We also note that all of the manipulations performed above are gauge-dependent and rely on the Regge-Wheeler gauge.
While it might be possible to work out the gauge-independent operator, we do not yet have a formal proof of this claim.
\subsection{Coupling Coefficients}\label{sec:nonlinear-coupling}

We now apply the operator formalism from the previous section to understand the evolution of the mode amplitudes beyond linear order.
The main point is that the mode expansion introduced in Sec.~\ref{sec:operator} can only be applied to a sourced perturbation whose source satisfies Eq.~\eqref{eq:constraints-on-source}.
We therefore first describe the constraints imposed on the tidal field at first order, before extending the construction to second order.

At first order, the metric perturbation outside the star can be fully described by a Zerilli--Moncrief function $Z_{\ell m}^{(1)}$ satisfying
\begin{align}
    \frac{\partial^2 Z_{\ell m}^{(1)}}{\partial r^{\ast}{}^{2}}
    +
    \left(
    \omega^2
    -
    V_{\ell}
    \right)
    Z_{\ell m}^{(1)}
    =
    0
    \,,
    \label{eqn:zerilli_eqn}
\end{align}
where $V_{\ell}$ is the Zerilli potential and
\begin{align}
    r^\ast
    =
    r
    +
    2M
    \ln
    \bigg(
    \frac{r}{2M}
    -
    1
    \bigg)
\end{align}
is the tortoise coordinate.
The source vanishes because the exterior spacetime is vacuum at first order.

In the region outside the star, we separate the homogeneous solution into a multipolar piece generated by the star and a tidal piece associated with the externally applied field,
\begin{align}
    Z_{\ell m}^{(1)}
    =
    Z_{I_{\ell m}}^{(1)}
    +
    Z_{T_{\ell m}}^{(1)}
    \,,
\end{align}
where
\begin{subequations}
\begin{align}
    Z_{I_{\ell m}}^{(1)}
    &=
    \frac{4\pi I_{\ell m}^{(1)}}{2\ell+1}
    z_I(r)
    \,,\\
    Z_{T_{\ell m}}^{(1)}
    &=
    \frac{4\pi d_{\ell m}^{(1)}}{2\ell+1}
    z_T(r)
    \,.
\end{align}
\end{subequations}
Here, $z_I(r)$ and $z_T(r)$ are the two independent solutions of Eq.~\eqref{eqn:zerilli_eqn}.
In the static limit, their expressions are
\begin{subequations}
\begin{align}
    z_I(r)&=-\frac{5r}{24M^5(3M+2r)} [3 \left(-6 M^3+3 M r^2+r^3\right)\times \nonumber \\
    \ln{}& (1{}-2M/r)+2 M (13 M^2+12 M r+3 r^2)],\\
    z_T(r)&=\frac{2 r (-6 M^3+3 M r^2+r^3)}{9 M+6 r}.
\end{align}
\end{subequations}
The stellar and tidal multipole moments are defined through
\begin{subequations}\label{eqn:first_moment}
\begin{align}
    I_{\ell m}^{(n)}
    &=
    I_L^{(n)}
    \mathscr{Y}^{\langle L\rangle}_{\ell m}
    \,,\\
    d_{\ell m}^{(n)}
    &=
    -
    \frac{1}{(2\ell-1)!!}
    \sum_{n=1}\mathcal{E}_L^{(n)}
    \mathscr{Y}^{\langle L\rangle}_{\ell m}
    \,,\label{eqn:tidal_moment_definition}
    \\
    \mathcal{E}^{(n)L}
    &=
    -
    \frac{4\pi \ell!}{2\ell+1}
    \sum_{m=-\ell}^{\ell}
    d_{\ell m}^{(n)}
    \mathscr{Y}^{\ast\langle L\rangle}_{\ell m}
    \,.
\end{align}
\end{subequations}
where $\mathscr{Y}_{\ell m}^{\langle L\rangle}$ is an STF spherical-harmonic tensor defined on $s_{AB}$.

Motivated by this exterior decomposition, we similarly separate the first-order perturbation inside the star into the induced stellar response and the tidal contribution,
\begin{align}
    \bm{y}^{(1)}
    =
    \bm{y}_I^{(1)}
    +
    \bm{y}_T^{(1)}
    \,,
\end{align}
with
\begin{align}
    \bm{y}_I^{(1)}
    =
    \sum_s
    \tilde{a}_s
    \bm{y}_s
    \,.
\end{align}
Here, $\bm{y}_I^{(1)}$ is the interior perturbation associated with the induced multipolar response $Z_I^{(1)}$, while $\bm{y}_T^{(1)}$ is an interior continuation of the externally applied tidal field $Z_T^{(1)}$.

Since the complete first-order perturbation satisfies the sourceless linearized equations,
\begin{subequations}
\begin{align}
    \bm{E}_{\alpha\beta}
    \left[
    \bm{y}^{(1)}
    \right]
    &=
    0
    \,,\\
    \bm{E}_{\alpha}
    \left[
    \bm{y}^{(1)}
    \right]
    &=
    0
    \,,
\end{align}
\end{subequations}
the equations governing the induced response can be written as
\begin{subequations}\label{eq:first-order-induced-equations}
\begin{align}
    \bm{E}_{\alpha\beta}
    \left[
    \bm{y}_I^{(1)}
    \right]
    &=
    -
    \bm{E}_{\alpha\beta}
    \left[
    \bm{y}_T^{(1)}
    \right]
    \,,\\
    \bm{E}_{\alpha}
    \left[
    \bm{y}_I^{(1)}
    \right]
    &=
    -
    \bm{E}_{\alpha}
    \left[
    \bm{y}_T^{(1)}
    \right]
    \,.
\end{align}
\end{subequations}
For $\bm{y}_I^{(1)}$ to admit the mode expansion above, the effective source on the right-hand side of Eq.~\eqref{eq:first-order-induced-equations} must satisfy Eq.~\eqref{eq:constraints-on-source}.
Using Eq.~\eqref{eq:linear-constraints}, this requires
\begin{subequations}\label{eq:first-order-tidal-constraints}
\begin{align}
    \bm{H}
    \left[
    \bm{y}_T^{(1)}
    \right]
    &=
    0
    \,,\\
    \bm{P}_{\rho}
    \left[
    \bm{y}_T^{(1)}
    \right]
    &=
    0
    \,.
\end{align}
\end{subequations}
Thus, the interior continuation of the first-order tidal field is required to satisfy the homogeneous Hamiltonian and momentum constraints.
It is not required to satisfy the complete sourceless evolution equations.
Any interior tidal field satisfying Eq.~\eqref{eq:first-order-tidal-constraints} and matching smoothly to the exterior tidal solution provides an admissible definition of $\bm{y}_T^{(1)}$.

Equation~\eqref{eq:first-order-induced-equations} can be used to define a mode $p$'s linear overlap integral~\cite{hegade26a,hegade26b}
\begin{align}
    \frac{4\pi}{2\ell + 1}d_{\ell m}^{(1)}I_p&\equiv -\frac{1}{MR^\ell}\int r^2 e^{\lambda/2}drd\Omega e^{\nu/2}
    \nonumber \\
    & \times 
    \bigg[\xi_{\alpha,p}\bm{E}^\alpha[\bm{y}_T^{(1)}] + \frac{h_{\alpha\beta,p}\bm{E}^{\alpha\beta}[\bm{y}_T^{(1)}]}{16\pi}\bigg],
\end{align}
which leads to the linear amplitude of the mode
\begin{align}
    \tilde{a}_p^{(1)}(\omega) = \frac{4\pi}{2\ell + 1}\frac{d^{(1)}_{\ell m}}{\mathscr{N}_p M/R^{\ell + 1}}\frac{I_p}{1-(\omega/\omega_p)^2}.
    \label{eqn:linear_amplitude}
\end{align}

The identification of $\bm{y}_I^{(1)}$ with the induced response is fixed by matching its metric perturbation to $Z_I^{(1)}$ at the stellar surface,
\begin{align}
    Z_{I_{\ell m}}^{(1)}(R)
    =
    \sum_s
    \tilde{a}_s^{(1)}
    Z_s(R)
    \,,
\end{align}
where $Z_s$ is the Zerilli--Moncrief function constructed solely from the eigenfunction of mode $s$ at the stellar surface.
In the Newtonian limit,
\begin{align}
    \lim_{c\rightarrow\infty}
    \frac{Z_s(R)}{z_I(R)}
    =
    \frac{4\pi}{2\ell+1}
    MR^\ell I_s
    \,,
\end{align}
where $I_s$ is the Newtonian overlap integral.

We now extend this construction to second order.
At second order, the perturbation equations can be written in the form
\begin{subequations}\label{eq:second-order-equations}
\begin{align}
    \bm{E}_{\alpha\beta}
    \left[
    \bm{y}^{(2)}
    \right]
    &=
    \bm{S}_{\alpha\beta}
    \left[
    \bm{y}^{(1)},
    \bm{y}^{(1)}
    \right]
    \,,\\
    \bm{E}_{\alpha}
    \left[
    \bm{y}^{(2)}
    \right]
    &=
    \bm{S}_{\alpha}
    \left[
    \bm{y}^{(1)},
    \bm{y}^{(1)}
    \right]
    \,.
\end{align}
\end{subequations}
Equation~\eqref{eq:second-order-equations} defines our convention for the second-order source.
The quantities
\begin{align}
    \bm{S}_{\alpha\beta}
    \left[
    \bm{y}_A,
    \bm{y}_B
    \right]
    \,,
    \qquad
    \bm{S}_{\alpha}
    \left[
    \bm{y}_A,
    \bm{y}_B
    \right]
\end{align}
denote the bilinear sources obtained from the terms quadratic in the lower-order perturbations when the exact Einstein and fluid equations are expanded about the TOV background.

At second order, we again separate the perturbation into an induced response and a tidal contribution,
\begin{align}
    \bm{y}^{(2)}
    =
    \bm{y}_I^{(2)}
    +
    \bm{y}_T^{(2)}
    \,.
\end{align}
The equations governing the induced response are therefore
\begin{subequations}\label{eq:second-order-induced-equations}
\begin{align}
    \bm{E}_{\alpha\beta}
    \left[
    \bm{y}_I^{(2)}
    \right]
    &=
    \bm{S}_{\alpha\beta}
    \left[
    \bm{y}^{(1)},
    \bm{y}^{(1)}
    \right]
    -
    \bm{E}_{\alpha\beta}
    \left[
    \bm{y}_T^{(2)}
    \right]
    \,,\\
    \bm{E}_{\alpha}
    \left[
    \bm{y}_I^{(2)}
    \right]
    &=
    \bm{S}_{\alpha}
    \left[
    \bm{y}^{(1)},
    \bm{y}^{(1)}
    \right]
    -
    \bm{E}_{\alpha}
    \left[
    \bm{y}_T^{(2)}
    \right]
    \,.
\end{align}
\end{subequations}

For $\bm{y}_I^{(2)}$ to be expanded in the stellar eigenmodes, the effective source on the right-hand side of Eq.~\eqref{eq:second-order-induced-equations} must again satisfy Eq.~\eqref{eq:constraints-on-source}.
Consequently, the second-order tidal field must satisfy
\begin{subequations}\label{eq:second-order-tidal-constraints}
\begin{align}
    \bm{H}
    \left[
    \bm{y}_T^{(2)}
    \right]
    &=
    \bm{S}_{\alpha\beta}
    \left[
    \bm{y}^{(1)},
    \bm{y}^{(1)}
    \right]
    u^\alpha u^\beta
    \,,\\
    (-i\omega)
    \bm{P}_{\rho}
    \left[
    \bm{y}_T^{(2)}
    \right]
    &=
    \bm{S}_{\alpha\beta}
    \left[
    \bm{y}^{(1)},
    \bm{y}^{(1)}
    \right]
    u^\alpha q^\beta{}_{\rho}
    \,.
\end{align}
\end{subequations}
The second-order tidal field therefore satisfies inhomogeneous Hamiltonian and momentum constraints, with sources determined by the quadratic first-order perturbations.
Once $\bm{y}_T^{(2)}$ is chosen to satisfy Eq.~\eqref{eq:second-order-tidal-constraints} together with the appropriate exterior matching conditions, the remaining second-order source satisfies Eq.~\eqref{eq:constraints-on-source} and can be projected consistently onto the stellar eigenmodes.

For later convenience, consider a contribution to the quadratic source generated by two first-order perturbations $\bm{y}_A$ and $\bm{y}_B$.
We denote the corresponding contribution to the second-order tidal field by
\begin{align}
    \bm{y}_T^{(2)}
    \left[
    \bm{y}_A,
    \bm{y}_B
    \right]
    \,,
\end{align}
and define the associated constraint-preserving source by
\begin{subequations}\label{eq:constraint-preserving-source}
\begin{align}
    \bm{\mathcal{S}}_{\alpha\beta}
    \left[
    \bm{y}_A,
    \bm{y}_B
    \right]
    &\equiv
    \bm{S}_{\alpha\beta}
    \left[
    \bm{y}_A,
    \bm{y}_B
    \right]
    -
    \bm{E}_{\alpha\beta}
    \left[
    \bm{y}_T^{(2)}
    \left[
    \bm{y}_A,
    \bm{y}_B
    \right]
    \right]
    \,,\\
    \bm{\mathcal{S}}_{\alpha}
    \left[
    \bm{y}_A,
    \bm{y}_B
    \right]
    &\equiv
    \bm{S}_{\alpha}
    \left[
    \bm{y}_A,
    \bm{y}_B
    \right]
    -
    \bm{E}_{\alpha}
    \left[
    \bm{y}_T^{(2)}
    \left[
    \bm{y}_A,
    \bm{y}_B
    \right]
    \right]
    \,.
\end{align}
\label{eqn:constraint_fixing_subtraction}
\end{subequations}
By construction,
\begin{align}
    \bm{\mathcal{S}}_{\alpha\beta}
    \left[
    \bm{y}_A,
    \bm{y}_B
    \right]
    u^\alpha u^\beta
    =
    0
    \,,
    \qquad
    \bm{\mathcal{S}}_{\alpha\beta}
    \left[
    \bm{y}_A,
    \bm{y}_B
    \right]
    u^\alpha q^\beta{}_{\rho}
    =
    0
    \,.\label{eqn:constraint_impose}
\end{align}
It is therefore $\bm{\mathcal{S}}$, rather than the bare quadratic source $\bm{S}$, that can be inserted directly into the operator projection derived in Sec.~\ref{sec:operator}.

With this in mind, we define the nonlinear coupling coefficients as
\begin{widetext}
\begin{subequations}
\begin{align}
    \kappa_{abc}
    &\equiv
    \frac{R}{M^2}
    \int
    \sqrt{-g}d^3x
    \bigg\{
    \xi_{\alpha,a}
    \bm{\mathcal{S}}^\alpha
    \left[
    \bm{y}_b,
    \bm{y}_c
    \right]
    +
    \frac{h_{\alpha\beta,a}}{16\pi}
    \bm{\mathcal{S}}^{\alpha\beta}
    \left[
    \bm{y}_b,
    \bm{y}_c
    \right]
    \bigg\}
    \,,\\
    \kappa_{ab\ell m}
    \hat{d}_{\ell m}
    &\equiv
    \frac{R}{2M^2}
    \int
    \sqrt{-g}d^3x
    \bigg\{
    \xi_{\alpha,a}
    \bm{\mathcal{S}}^\alpha
    \left[
    \bm{y}_b,
    \bm{y}_{T_{\ell m}}
    \right]
    +
    \frac{h_{\alpha\beta,a}}{16\pi}
    \bm{\mathcal{S}}^{\alpha\beta}
    \left[
    \bm{y}_b,
    \bm{y}_{T_{\ell m}}
    \right]
    \bigg\}
    \,,\\
    \kappa_{a\ell m\ell' m'}
    \hat{d}_{\ell m}
    \hat{d}_{\ell' m'}
    &\equiv
    \frac{R}{M^2}
    \int
    \sqrt{-g}d^3x
    \bigg\{
    \xi_{\alpha,a}
    \bm{\mathcal{S}}^\alpha
    \left[
    \bm{y}_{T_{\ell m}},
    \bm{y}_{T_{\ell' m'}}
    \right]
    +
    \frac{h_{\alpha\beta,a}}{16\pi}
    \bm{\mathcal{S}}^{\alpha\beta}
    \left[
    \bm{y}_{T_{\ell m}},
    \bm{y}_{T_{\ell' m'}}
    \right]
    \bigg\}
    \,.
\end{align}
\end{subequations}

With the above definitions, the lowest-order nonlinear amplitude of mode $p$ takes the form
\begin{align}
    \frac{1}{2}
    \mathscr{N}_p
    \left(
    1
    -
    \omega^2/\omega_p^2
    \right)
    \tilde{a}_p^{(2)}(\omega)
    &=
    \sum_{bc}
    \kappa_{pbc}
    \big[\tilde{a}_b^{(1)}
    \ast 
    \tilde{a}_c^{(1)}\big]
    +
    \sum_{b\ell m}
    \kappa_{pb\ell m}
    \big[
    \tilde{a}_b^{(1)}
    \ast
    \hat{d}_{\ell m}^{(1)}
    \big]
    +
    \sum_{\ell m\ell' m'}
    \kappa_{p\ell m\ell' m'}
    \big[
    \hat{d}_{\ell m}^{(1)}
    \ast
    \hat{d}_{\ell' m'}^{(1)}
    \big]
    +
    \mathcal{O}
    \left(
    \hat{d}_{\ell m}^3
    \right)
    \label{eqn:second_amplitude}
    \,.
\end{align}
where
\begin{align}
    \hat{d}_{\ell m}^{(1)}=\frac{4\pi}{2\ell + 1}\frac{R^{\ell + 1}}{M}d_{\ell m}^{(1)}
    ,&& [f\ast g](\omega)\equiv \frac{1}{2\pi}\int f(\omega')g(\omega-\omega')\,d\omega'.
\end{align}
\end{widetext}
The three coupling coefficients appearing above have a simple physical interpretation.
The coefficient $\kappa_{abc}$ describes the nonlinear coupling between stellar modes.
Modes $b$ and $c$ generate a quadratic source that drives mode $a$.
The coefficient $\kappa_{ab\ell m}$ describes the nonlinear coupling between a stellar mode and the external tidal field.
In this case, the quadratic source is generated by mode $b$ together with the $(\ell,m)$ component of the first-order tidal perturbation, and this source drives mode $a$.
Finally, $\kappa_{a\ell m\ell' m'}$ describes the direct quadratic driving of mode $a$ by two components of the external tidal field.
The corresponding source is generated by the interaction between the $(\ell,m)$ and $(\ell',m')$ tidal perturbations.
Unlike $\kappa_{abc}$ and $\kappa_{ab\ell m}$, this contribution does not require a stellar mode to be excited at first order.

Schematically, the three classes of couplings can therefore be summarized as
\begin{subequations}
\label{eqn:coupling_channel}
\begin{align}
    \kappa_{abc}
    &: \qquad
    \text{mode}
    \times
    \text{mode}
    \rightarrow
    \text{mode}
    \,,
    \\
    \kappa_{ab\ell m}
    &: \qquad
    \text{mode}
    \times
    \text{tide}
    \rightarrow
    \text{mode}
    \,,
    \\
    \kappa_{a\ell m\ell' m'}
    &: \qquad
    \text{tide}
    \times
    \text{tide}
    \rightarrow
    \text{mode}
    \,.
\end{align}
\end{subequations}
In each case, the first index labels the mode onto which the quadratic source is projected, while the remaining indices label the two first-order perturbations that generate the source.

In Newtonian gravity, the externally applied gravitational potential enters the perturbation equations linearly and does not gravitationally self-interact.
In contrast, in general relativity, the first-order tidal metric perturbations themselves source the second-order Einstein equations.
Consequently, even in the absence of a first-order stellar mode, two tidal perturbations can generate a second-order gravitational source that excites the star.

The summations in the first two terms on the right-hand side of Eq.~\eqref{eqn:second_amplitude} run over all modes, which includes the driven mode $p$ itself.
When these self-coupling terms are moved to the left-hand side, we can see that they induce a frequency shift that depends on the mode's amplitude. 
For $f$-mode tide, the shift lowers the frequency closer to the resonance~\cite{yu23, pitre25}.

\section{Example Application: Static Quadratic Love Number from Mode Decomposition\label{sec:p2}}

We now apply the nonlinear mode-coupling framework to a star's static, second-order tidal response. We begin by defining the response function for a general induced multipole $(\ell,m)$. We then express that multipole in terms of the mode amplitudes and coupling coefficients derived in Sec.~\ref{sec:nonlinear-coupling}. Finally, we specialize to the quadrupolar $(\ell,m)=(2,0)$ response and use the $f$-mode for a numerical demonstration.

\subsection{Quadratic Tidal Response}

Consider a star subjected to a weak external tidal field generated by its binary companion. At linear order, the induced multipole moment is proportional to the applied tidal moment:
\begin{align}
    I_{\ell m}^{(1)}(\omega)
    =2k_{\ell}(\omega)R^{2\ell+1}d_{\ell m}^{(1)}(\omega),
    \label{eqn:linear_response_function}
\end{align}
where $K_{\ell}(\omega)$ is the linear response function and $d_{\ell m}^{(1)}$ is defined in Sec.~\ref{sec:second_order_perturbation_theory}. The zero-frequency limit of Eq.~\eqref{eqn:linear_response_function} describes the static linear response.

At second order, the induced response receives contributions from all three coupling channels in Eq.~\eqref{eqn:coupling_channel}: interactions between two first-order induced perturbations, between an induced perturbation and an external tide, and between two external tides. 
Since the first-order induced perturbation is itself linear in the applied tidal field, each channel contributes at quadratic order in the tidal field.

Thus, consider a pair of first-order tidal multipoles with ranks $\ell_1$ and $\ell_2$. 
We denote the irreducible rank-$\ell$ part of their product by $(\mathcal{E}_1^{(1)}\mathcal{E}_2^{(1)})_L$ and define the second-order tidal moment
We then define the quadratic response function $p_{\ell m}$ by
\begin{align}
    I_{\ell m}^{(2)}(\omega)
   & \equiv -\frac{4}{(2\ell-1)!!}p_{\ell m}(\omega)
    \frac{R^{\ell+\ell_1+\ell_2+2}}{M}
    \nonumber \\
    &\quad \times \big(\mathcal{E}_1^{(1)}\mathcal{E}_2^{(1)}\big)_L(\omega)
    \mathscr{Y}_{\ell m}^{\langle L\rangle }.
    \label{eqn:second_response}
\end{align}
In principle, the second-order correction to the tidal field $\mathcal{E}^{(2)}_L$ perturbs the star's multipole at the same order.
We use the freedom to renormalize the tidal field to choose 
\begin{align}
    \mathcal{E}^{(2)}_L=0.
\end{align}

Now, the task is to compute $I_{\ell m}^{(2)}$ from the mode decomposition. We do so using the coupling coefficients and overlap integrals introduced in Sec.~\ref{sec:nonlinear-coupling}.

\subsection{Eigenmode Decomposition of Static Quadratic Tidal Response}

We proceed in the same manner as we derived the linear response function using modes in Paper~\citetalias{hegade26b}.
We use the fact that the second-order perturbation in the star's exterior can be described in terms of the solution to the sourced Zerilli--Moncrief equation,
\begin{align}
    \frac{\partial^2 Z_{\ell m}^{(2)}}{\partial r^{\ast}{}^{2}}
    +
    \left(
    \omega^2
    -
    V_{\ell}
    \right)
    Z_{\ell m}^{(2)}
    =
    S_{\ell m}^{(2)}
    \,,
    \label{eqn:sourced_zerilli_eqn}
\end{align}
where $S_{\ell m}^{(2)}$ is quadratic in the first-order perturbations. 
This term can be constructed using $\bm{S}^{\alpha\beta}$ in Eq.~\eqref{eq:schematic-perturbation-equations}.
We refer readers to Ref.~\cite{martel04} for the details on constructing $S_{\ell m}^{(2)}$.
We then solve Eq.~\eqref{eqn:sourced_zerilli_eqn} in the static limit, $\omega=0$.

We separate the resulting solution into induced, tidal, and coupled pieces:
\begin{align}
    Z_{\ell m}^{(2)}
    =Z_{I_{\ell m}}^{(2)}+Z_{T_{\ell m}}^{(2)}+Z_{C_{\ell m}}^{(2)}.
\end{align}
As at linear order, $Z_{I_{\ell m}}^{(2)}$ and $Z_{T_{\ell m}}^{(2)}$ solve the homogeneous Zerilli--Moncrief equation.
The coupled piece $Z_{C_{\ell m}}^{(2)}$ is new, and it corresponds to the particular solution due to the source.
It is defined up to the addition and subtraction of the homogeneous solutions. 
We use this property to isolate the multipolar contribution to the homogeneous solutions as
\begin{subequations}
\begin{align}
    Z_{I_{\ell m}}^{(2)}&=\frac{4\pi I_{\ell m}^{(2)}}{2\ell + 1}z_I(r),
    \label{eqn:zI2}\\
    Z_{T_{\ell m}}^{(2)}&=\frac{4\pi d_{\ell m}^{(2)}}{2\ell + 1}z_T(r)=0.
\end{align}
\end{subequations}
Finally, the coupled piece $Z_{C_{\ell m}}^{(2)}$ includes terms quadratic in the first-order multipole moments $d_{\ell m}^{(1)}$ and $I_{\ell m}^{(1)}$.
That is, 
\begin{align}
    Z_{C_{\ell m}}^{(2)}=Z_{C_{\ell m},II}^{(2)}+Z_{C_{\ell m},IT}^{(2)}+Z_{C_{\ell m},TT}^{(2)}
\end{align}
where $II$, $IT$, and $TT$ indicates the coupling across induced and tidal multipoles.
We provide the full expression of $Z_{C_{\ell m}}^{(2)}$ in a supplementary \textsc{Mathematica} notebook.

We now relate the interior mode expansion to the induced
multipole appearing in the exterior solution.
Equations~\eqref{eqn:constraint_fixing_subtraction} and \eqref{eqn:constraint_impose} ensure that the source driving
$\bm{y}_I^{(2)}$ satisfies the constraints required for
the mode projection.
Therefore, we use the formal expansion in the static limit
\begin{align}
    \bm{y}_I^{(2)}
    =\sum_s \tilde a_s^{(2)}(0)\bm{y}_s,
\end{align}
where the amplitudes are determined by Eq.~\eqref{eqn:second_amplitude} with $\omega = 0$. 
Also, note that the contribution to the Zerilli--Moncrief function that is linear in $\bm{y}_I^{(2)}$ represents the same degree of freedom as $\bm{y}_I^{(1)}$ for linear perturbation.
This means we have~\cite{hegade26b}
\begin{align}
    Z_{I_{\ell m},{\rm int}}^{(2)}(R)
    =
    \sum_{\substack{s\\ \ell_s=\ell\\
    m_s=m}}
    \tilde a_s^{(2)}(0)Z_s(R).
\end{align}

{The complete interior expression entering the matching
conditions also contains contributions determined by
the first-order perturbations. These include the
constraint-fixing field $\bm{y}_T^{(2)}[\bm{y}_A, \bm{y}_B]$ introduced
in Eqs.~\eqref{eqn:constraint_fixing_subtraction}, together with the quadratic terms required
when reconstructing and evaluating the perturbation
at the stellar surface.
We denote their combined
contribution by $Z_{C_{\ell m},{\rm int}}^{(2)}$, which also splits into
\begin{align}
Z^{(2)}_{C_{\ell m},{\rm int}}
=
Z^{(2)}_{C_{\ell m},II,{\rm int}}
+
Z^{(2)}_{C_{\ell m},IT,{\rm int}}
+
Z^{(2)}_{C_{\ell m},TT,{\rm int}.}
\end{align}}

Expressing
the surface matching relation at $r=R$, we have 
\begin{align}
    \frac{4\pi I_{\ell m}^{(2)}}{2\ell+1}
    z_I(R)
    +Z_{C_{\ell m},II}^{(2)}(R)
    &=
    \sum_{\substack{s
    \\
    \ell_s=\ell
    \\
    m_s=m}}
    \tilde a_s^{(2)}(0)Z_s(R)
    \nonumber \\
    &\quad + Z_{C_{\ell m},II,{\rm int}}^{(2)}(R).
\end{align}
Now further define $Z_{st\ell m}$ such that 
\begin{align}
    \sum_{st}\tilde{a}_s(0)\tilde{a}_t(0)Z_{st\ell m}(R)\equiv Z^{(2)}_{C_{\ell m},II,{\rm int}}(R) - Z^{(2)}_{C_{\ell m},II}(R).
\end{align}
Here $Z_{ab\ell m}$ is the part of the Zerilli--Moncrief function that is bilinear in modes $a$ and $b$, obtained from the second-order metric perturbation they source with the corresponding exterior particular solution subtracted; in the Newtonian limit it reduces to the quadratic overlap $J_{ab\ell m}$~\cite{w12}. We provide its explicit expression in the supplementary notebook.

Consequently, we can write the multipole moment as 
\begin{align}
    I_{\ell m}^{(2)}
    &=
    \frac{2\ell+1}{4\pi z_I(R)}
    \bigg[
    \sum_{\substack{s
    \\
    \ell_s=\ell
    \\m_s=m}}
    \tilde a_s^{(2)}(0)Z_s(R)
    \nonumber \\
    &\quad 
    + \sum_{st}
    \tilde{a}_s^{(1)}(0)\tilde{a}_t^{(1)}(0)
    Z_{st\ell m}(R)
    \bigg].\label{eqn:I2}
\end{align}

Substituting Eq.~\eqref{eqn:I2} into the definition of $p_{\ell m}$ in Eq.~\eqref{eqn:second_response}, 
and using Eqs.~\eqref{eqn:linear_amplitude} and \eqref{eqn:second_amplitude} for modes' amplitudes,
we obtain the static quadratic response function, 
for modes and tidal multipoles of the same $\ell$ as the induced multipole
\begin{widetext}
\begin{align}
    p_{\ell m}&=\frac{4\pi}{2\ell + 1}\sum_{\substack{abc\\\ell_a =\ell\\m_a=m}}\frac{\kappa_{abc}I_bI_c}{\mathscr{N}_a\mathscr{N}_b\mathscr{N}_c}\frac{Z_a/z_I }{MR^{\ell}}\frac{d_{b}^{(1)}d_{c}^{(1)}}{d_{\ell m}^{(2)}}
    +\frac{4\pi}{2\ell + 1}\sum_{\substack{ab\ell' m'\\\ell_a=\ell\\m_a=m}}\frac{\kappa_{ab{\ell' m'}}I_b}{\mathscr{N}_{a}\mathscr{N}_{b}}\frac{Z_a/z_I}{M R^{\ell}}\frac{d_b^{(1)}d^{(1)}_{{\ell' m'}}}{d^{(2)}_{\ell m}}\nonumber\\
    &\hphantom{=}{}+\frac{2\pi}{2\ell + 1}\sum_{ab}\frac{I_a I_b}{\mathscr{N}_a\mathscr{N}_b}\frac{Z_{ab{\ell m}}/z_I}{MR^{\ell}}\frac{d_a^{(1)}d_b^{(1)}}{d_{\ell m}^{(2)}}+\frac{4\pi}{2\ell + 1}\sum_{\substack{a\ell' m' \ell'' m''\\\ell_a=\ell\\m_a=m}}\frac{\kappa_{a {\ell' m'}{\ell'' m''}}}{\mathscr{N}_a }\frac{Z_a/z_I}{MR^{\ell}}\frac{d_{{\ell' m'}}^{(1)}d_{{\ell''m''}}^{(1)}}{d_{\ell m}^{(2)}} .
    \label{eqn:p2_full}
\end{align}

We now specialize this expression to $(\ell,m)=(2,0)$ of the induced multipole,
retaining only the $f$-modes in the sums. This gives
\begin{align}
    p_{20}=-\sqrt{\frac{\pi}{5}}\frac{\kappa_{fff}I_f^2}{\mathscr{N}_f^3}\frac{Z_f/z_I}{MR^2}-\sqrt{\frac{\pi}{5}}\frac{\kappa_{ff{20}}I_f}{\mathscr{N}_f^2}\frac{Z_f/z_I}{MR^2}-\frac{1}{2}\sqrt{\frac{\pi}{5}}\frac{I_f^2}{\mathscr{N}_f^2}\frac{Z_{ff{20}}/z_I}{MR^2}-\sqrt{\frac{\pi}{5}}\frac{\kappa_{f{20}{20}}}{\mathscr{N}_f}\frac{Z_f/z_I}{MR^2} \label{eqn:p2_fmodes}
\end{align}
\end{widetext}
where $Z_f=Z_f(R)$, $z_I=z_I(R)$, and $m_f=0$ in the coupling integral. We have used the relation between the first- and second-order $m=0$ tidal moments,
\begin{align}
    d_{20}^{(2)}=-\bigg(\frac{16\pi}{5}\bigg)^{1/2}
    \Big[d_{20}^{(1)2}
    -2d_{22}^{(1)}d_{2,-2}^{(1)}\Big],
\end{align}
as derived in Appendix~\ref{app:second_order_tidal_moments}.

The Newtonian limit provides a further comparison. At leading order, Eq.~\eqref{eqn:p2_fmodes} becomes\footnote{The last term in Eq.~\eqref{eqn:p2_fmodes} does not have any term at order $\mathcal{O}(c^{0})$.}
\begin{align}
    p_{20}=-\frac{4\pi}{5}\sqrt{\frac{\pi}{5}}\frac{\kappa_{fff}I_f^3}{\mathscr{N}_f^3}
    -\frac{6\pi}{5}\sqrt{\frac{\pi}{5}}\frac{J_{ff20}I_f^2}{\mathscr{N}_f^2};
\end{align}
\emph{cf}. Eq.~(3) of Ref.~\cite{yu26}.

\subsection{Provisional Numerical Comparison}

We next evaluate $p_{20}$ in the static limit using Eq.~\eqref{eqn:p2_fmodes}.
We construct neutron-star models of masses $M=1.4\,M_\odot$ and $2.0\,M_\odot$ by solving the TOV equations with the SLy4 equation of state~\cite{chabanat98}.
We calculate their $f$-modes following the procedure described in Paper~\citetalias{hegade26b}.

When evaluating the coupling coefficients as described in Sec.~\ref{sec:nonlinear-coupling}, we omit the second term on the right-hand side of Eq.~\eqref{eq:constraint-preserving-source}. 
The resulting second-order equation of motion lacks a constraint-fixing piece, which, however, vanishes at the Newtonian order~(see Appendix~\ref{app:constraint}).
The dominant contribution to $p_{20}$ comes from the pure-fluid sector at Newtonian order. 

\begingroup
\renewcommand{\arraystretch}{1.2}
\begin{table}[h]
\caption{\label{tab:p2_numbers}
Numerical comparison of $p_{20}(\omega=0)$ computed using metric matching~\cite{pitre25} and the $f$-mode approximation for neutron stars of masses $1.4\,M_\odot$ and $2.0\,M_\odot$ constructed using the SLy4 equation of state. 
The comparison is preliminary, as we have omitted the terms required to impose the Hamiltonian and momentum constraints for the second-order tidal field.
We show that the correction has a vanishing contribution at Newtonian order in Appendix~\ref{app:constraint}.
}
\begin{ruledtabular}
\begin{tabular}{cccc}
\textrm{$M$ [$M_\odot$]}
&
\textrm{$f$-mode sum}
&
\textrm{Direct matching}
&
\textrm{$\%$ difference}
\\
\colrule
$1.4$ & $6.7\times 10^{-2}$ & $6.4\times 10^{-2}$& $4.7\%$
\\
$2.0$ & $8.6\times 10^{-3}$ & $8.3\times 10^{-3}$ & $3.6\%$
\end{tabular}
\end{ruledtabular}
\end{table}
\endgroup

Table~\ref{tab:p2_numbers} compares this calculation with the result obtained without mode decomposition~\cite{pitre25}, in which the external perturbation is continued into the stellar interior. The two results differ by approximately $4\%$.

Two features of the present calculation may contribute to this difference. First, the mode expansion uses a non-positive-definite orthogonality operator. Convergence of the resulting mode sum is therefore not guaranteed, and we treat it as a formal expansion. Second, the calculation omits the constraint-fixing term in Eq.~\eqref{eq:constraint-preserving-source}.
We will perform a more detailed numerical analysis accounting for this term in future work~\cite{nonlinear-f-modes-gr}.

\section{Conclusion\label{sec:conclusion}}
In this work, we developed a general, relativistic Lagrangian theory framework for describing dynamical nonlinear perturbations of non-rotating, spherically symmetric stars. 
Explicitly, we derived the adiabatic Lagrangian variations of the fluid variables through third order. 
We then specialized the formalism to polar perturbations in the Regge-Wheeler gauge. 
We reduced the sourced field equations to a system of a set of sourced master equations, keeping their source functions general, and explicitly evaluated the perturbations entering these sources through second order.
By combining this construction with the relativistic operator formalism, we projected the nonlinear sources onto the stellar eigenmodes, which defined coupling coefficients describing interactions among internal modes and the externally applied tidal field.
The resulting amplitude equations retain the familiar driven-oscillator structure at nonlinear level seen in Newtonian theory~\cite{vhoolst93}.

As an initial application and validation of the framework, we expressed the static, nonlinear quadratic tidal response in terms of the relativistic mode overlaps, coupling coefficients, and exterior metric perturbations. The resulting expression reduces to the known Newtonian mode-sum formula in the weak-field limit.
We also calculated the value of the static nonlinear Love number using a partial set of terms in the mode-coupling to show that the formalism is capable of reproducing the known results to approximately 4\% accuracy, subject to correction due to the constraint fixing term at the first PN order.
We will perform a more detailed numerical analysis in our future work~\cite{nonlinear-f-modes-gr}, in which we will systematically investigate the impact of dynamical nonlinearity over a wide range of equation of state.
Also, we will move beyond both the mode-approximation and the approximations used here to obtain the dynamical nonlinearity to second order in perturbation theory~\cite{nonlinear_dynamical_love_numbers}.

The present construction is restricted to perturbations of a non-rotating spherical background and is formulated in the Regge--Wheeler gauge. Moreover, the relativistic mode expansion should be regarded as a formal representation rather than a convergent spectral decomposition.
Subject to these approximations, the framework provides a practical relativistic counterpart to the mode-coupling methods widely used in Newtonian stellar dynamics.
Our framework is fully dynamical and can be applied to frequency-dependent nonlinear tides, including nonlinear shifts of mode frequencies and resonant mode interactions.
Evaluating these effects for realistic neutron-star models will allow the effect of hydrodynamic nonlinearities on the late inspiral and the resulting gravitational-wave signal to be assessed.
\section*{Acknowledgment}
K.J.K. and T.V. acknowledge support from NSF grants 2012086 and 2309360, the Alfred P. Sloan Foundation through grant number FG-2023-20470, and the BSF through award number 2022136.
H.Y. acknowledges support by NSF award No. PHY-2308415, CAREER award No. PHY-254157, and Montana NASA EPSCoR Research Infrastructure Development under award No. 80NSSC22M0042.
N.Y.~acknowledges support from the Simons Foundation through Award No. 896696, the Simons Foundation International through Award No. SFI-MPS-BH-00012593-01, the NSF through Grant No.
PHY-25-12423.
\input{appendix}
\bibliography{ref}
\end{document}

%% file: appendix.tex
\appendix
\section{Lagrangian Variation of Metric Determinant}\label{appendix:lagrangian-variation}
The Lagrangian change in the metric determinant can be obtained using the following identity
\begin{align}
    \frac{\Delta g }{g}
    &=
    \frac{\chi_{\param}^* g_{s} 
    -
    g}{g}
    = 
    \frac{\mathrm{Det} \left[ \chi_{\param}^* g_{\param,\alpha \beta} \right]
    -
    \mathrm{Det}[g_{\alpha \beta}]}
    {\mathrm{Det}[g_{\alpha \beta}]}
    \nonumber\\
    &=
    \frac{\mathrm{Det} \left[ \Delta g_{\alpha \beta} + g_{\alpha \beta} \right]
    -
    \mathrm{Det}[g_{\alpha \beta}]}
    {\mathrm{Det}[g_{\alpha \beta}]}
    \nonumber\\
    &=
    \mathrm{Det} \left[ \delta^{\alpha}_{\beta} + g^{\alpha \sigma} \Delta g_{\sigma \beta} \right]
    -
    1
    \,.
\end{align}
We can now expand the right-hand side in $\Delta g_{\sigma \beta}$ to arbitrary order using the Jacobi identity.
We display the expansion to $\mathcal{O}(\Delta g)^2$ here
\begin{align}\label{eqn:det_pert_0}
    \frac{\Delta g}{g}
    &=
    g^{\alpha\beta}\Delta g_{\alpha\beta}+
    \frac{1}{2}
    \Big[
    (g^{\alpha\beta}
    \Delta g_{\alpha\beta})^2
    -
    g^{\alpha\gamma}g^{\beta\delta}
    \Delta g_{\alpha\beta}\Delta g_{\gamma\delta}
    \Big]
    \nonumber\\
    &\quad 
    +
    \frac{1}{6}\Big[\big(g^{\alpha\beta}\Delta g_{\alpha\beta}\big)^3 
    -
    3 g^{\alpha\beta}\Delta g_{\alpha\beta} g^{\gamma\epsilon}g^{\delta\zeta}  \Delta g_{\gamma\delta}\Delta g_{\epsilon\zeta}
    \nonumber\\
    &\quad + 2g^{\alpha\gamma}g^{\beta\epsilon}g^{\delta\zeta} \Delta g_{\alpha\beta}\Delta g_{\gamma\delta}\Delta g_{\zeta\epsilon}\Big] + \mathcal{O}(\Delta g^4)
    \,.
\end{align}
We can simplify this further by using
\begin{align}
    \big(
    g_{\alpha \beta}
    +
    \Delta g_{\alpha \beta}
    \big)
    \big(
    g^{\beta \gamma}
    +
    \Delta g^{\beta \gamma}
    \big)
    =
    \delta_{\alpha}^{\gamma}
    \,,
\end{align}
which follows from the fact that the pullback of the inverse metric is the inverse
of the pulled-back metric.
Expanding this expression gives
\begin{align}
    g_{\alpha\beta}\Delta g^{\beta\gamma}
    +
    \Delta g_{\alpha\beta}g^{\beta\gamma}
    +
    \Delta g_{\alpha\beta}\Delta g^{\beta\gamma}
    =
    0
    \,.
\end{align}
Solving perturbatively for the  change in the inverse metric, we obtain
\begin{align}
    \Delta g^{\alpha\beta}
    &=
    -
    g^{\alpha\gamma}g^{\beta\delta}
    \Delta g_{\gamma\delta}
    +
    g^{\alpha\gamma}
    g^{\beta\epsilon}
    g^{\delta\zeta}
    \Delta g_{\gamma\delta}
    \Delta g_{\epsilon\zeta}
    \nonumber\\
    &
    -g^{\alpha \eta}g^{\beta\theta}g^{\gamma\epsilon}g^{\delta\zeta}\Delta g_{\epsilon \zeta}\Delta g_{\eta \gamma}\Delta g_{\theta \delta}+
    \mathcal{O}(\Delta g^4)
    \,.
\end{align}
Substituting in Eq.~\eqref{eqn:det_pert_0} we get
\begin{align}\label{eqn:det_pert}
    &\frac{\Delta g}{g}
    =
    g^{\alpha\beta}\Delta g_{\alpha\beta}+
    \frac{1}{2}
    \big[
    \big(g^{\alpha\beta}
    \Delta g_{\alpha\beta}\big)^2
    +
    \Delta g_{\alpha\beta}\Delta g^{\alpha\beta}
    \big]
    \nonumber\\
    &
    +
    \frac{1}{6}
    \big[\big(g^{\alpha\beta}\Delta g_{\alpha\beta}\big)^3 
    +
    3g^{\alpha\beta}\Delta g_{\alpha\beta}\Delta g_{\gamma\delta}\Delta g^{\gamma\delta}
    \nonumber \\
    &-g^{\alpha\gamma}g^{\beta\epsilon}g^{\delta\zeta}\Delta g_{\alpha\beta}\Delta g_{\gamma\delta}\Delta g_{\epsilon\zeta} 
    \big]
    +\mathcal{O}(\Delta g^4)\,.
\end{align}
Finally, we can relate $\Delta g/g$ to $\Delta \sqrt{|g|}/\sqrt{|g|}$.
Since the metric signature is unchanged under the perturbation,
\begin{align}
    \frac{\Delta \sqrt{|g|}}{\sqrt{|g|}}
    &=
    \sqrt{1+\frac{\Delta g}{g}}
    -
    1
    \,.
\end{align}
Using the generalized binomial expansion
\begin{align}
    (1+x)^{1/2}
    =
    1
    +
    \sum_{n=1}^{\infty}
    \binom{1/2}{n}
    x^n
\end{align}
together with
\begin{align}
    \binom{1/2}{n}
    =
    \frac{(-1)^{n-1}(2n-3)!!}{2^n n!}
    \,,
\end{align}
we obtain
\begin{align}
    \frac{\Delta \sqrt{|g|}}{\sqrt{|g|}}
    =
    \sum_{n=1}^{\infty}
    \frac{(-1)^{n-1}(2n-3)!!}{2^n n!}
    \left(
    \frac{\Delta g}{g}
    \right)^n
    \,.
\end{align}
The explicit expression to $\mathcal{O}(\Delta^3)$ is
\begin{align}\label{eq:variation-sqrt-det}
    \frac{\Delta \sqrt{|g|}}{\sqrt{|g|}}
    = 
    \frac{1}{2}\frac{\Delta g}{g} 
    -
    \frac{1}{8}\bigg(\frac{\Delta g}{g}\bigg)^2 
    +
    \frac{1}{16}\bigg(\frac{\Delta g}{g}\bigg)^3
    +
    \mathcal{O}(\Delta g^4)
\end{align}

We can repeat the derivation outlined above to get $\Delta \sqrt{q}/\sqrt{q}$ needed to evaluate Eq.~\eqref{eqn:volume_pert}; the explicit expression to $\mathcal{O}(\Delta q^3)$ is obtained by replacing $g \to  q$ and $\Delta g \to \Delta q$ in Eq.~\eqref{eq:variation-sqrt-det}.

\section{Spherical Harmonics Integral\label{appendix:spherical_harmonics_integral}}
Below lists the definition of the integrals involving three spherical harmonics uesd in the text:
\begin{subequations}
\begin{align}
    T_{abc}
    &
    \equiv 
    \int Y_a^\ast Y_b Y_c\, d\Omega,
    \\
    F_a 
    &\equiv 
    \int Y_a^\ast D_A Y_b D^A Y_c\, d\Omega, \\
    G_a 
    &\equiv 
    \int D_A D_B Y_a^\ast D_A Y_b D^B Y_c\, d\Omega,\\
    U_a 
    &\equiv 
    \int Y_a^\ast D_A D_B Y_b D^A D^B Y_c\, d\Omega.
\end{align}
\end{subequations}
The integrals can be analytically evaluated using the Gaunt coefficient and integration by parts~\cite{wu01}.

\section{Master Equation\label{appendix:master_equation}}
The master equation reads
\begin{widetext}
\begin{subequations}
\begin{align}
    K_a'
    &
    =- \frac{2 H_{0 a}{}}{r} + \frac{K_{a}{} \bigl(-3 + e^{\lambda } - 2 \ell_{a}{} + 8 e^{\lambda } \pi p r^2\bigr)}{2 r} -  \frac{8 e^{\lambda/2  } \pi \rho h W_{a}{}}{r} + \frac{i \ell_{a}{} (1 + \ell_{a}{}) a_{a}{} H_{1 a}{}}{2 r^2 \dot{a}_a}
    +
    \frac{S_{tr}}{r^{\ell_a}\dot{a}_a}
    \\
    W_a'
    &
    =
    e^{\lambda/2  }r(K_a-H_{0a}) -  \frac{e^{\lambda/2  } \ell_{a}{} (1 + \ell_{a}{}) V_{a}{}}{r} -  \frac{(1 + \ell_{a}{}) W_{a}{}}{r} -  \frac{e^{\lambda/2 } \rho h r X_{a}{}}{e^{ \nu/2 } \Gamma_{1}{} p}
    +\frac{2e^{\lambda/2}r^{1-\ell_a}S_{\Omega^2}}{(\ell_a-1)\ell_a (\ell_a + 1)(\ell_a+2)}
    \nonumber\\
    &\quad 
    - 
    \frac{e^{\lambda/2}r^{1-\ell_a}S_\Omega}{\ell_a(\ell_a+1)a_a p\Gamma_1}
    -
    \frac{e^{\lambda/2}r^{1-\ell_a}S_t}{\rho h \dot{a}_a}
    \\
    H_{1a}'
    &
    =
    -
    \frac{H_{1 a}{} \big(\ell_{a}{} + 2 e^{\lambda } (M/r) + 4 e^{\lambda } \pi  r^2 (p- \varepsilon) \big)}{r} + \frac{2i e^{\lambda } H_{0 a}{} \dot{a}_a}{a_{a}{}} -  \frac{i e^{\lambda } K_{a}{} \dot{a}_a}{a_{a}{}} + \frac{16i e^{\lambda } \pi \rho h V_{a}{} \dot{a}_a}{a_{a}{}} 
    \nonumber
    \\
    &\quad
    -\frac{2i r^{1-\ell_a}\big[ 2 S_{\Omega^2}\dot{a}_a-(-2+\ell_a+\ell_a^2)a_a S_{t\Omega} \big]}{(\ell_a-1)\ell_a(\ell_a + 1)(\ell_a+2)a_a^2 (-2M+r)}
    \\
    X_a'&
    =
    - \frac{i a_{a}{}  \big[e^{\nu } \ell_{a}{} (1 + \ell_{a}{}) - 2r^2  (\dot{a}_a/a_a)^2\big]}{4 e^{1/2 \nu } r^2 \dot{a}_a}
    H_{1 a}{}
    -
     \bigg[\frac{\ell_{a}{}}{r} +  (\ln \rho h)'\bigg] X_{a}{}
    -
    \frac{e^{\nu/2 } \ell_{a}{} (1 + \ell_{a}{}) {} p'}{\rho h r^2}V_{a}
    + \frac{e^{\nu/2 } {} \bigl(\rho h + r p'\bigr)}{\rho h r} 
    H_{0 a}
    \nonumber \\
    &\quad
    + 
    \frac{-2 e^{\lambda } \rho h \Gamma_{1}{} p r \ddot{a}_a + e^{\nu } a_{a} \big\{8 e^{\lambda } \pi \big(\rho h\big)^2 \Gamma_{1}{} p r + r \rho hp' \nu' + \Gamma_{1}{} p \big[p' \big(-4 -  r\lambda'+ r \nu'\big) + 2 r p''\big]\big\}}{2 e^{(\lambda+\nu)/2} a_{a}{} \rho h \Gamma_{1}{} p r^2}W_{a}
    \nonumber \\
    &\quad
    + \frac{e^{\nu/2 } {} \big[- \rho h \bigl(-3 + e^{\lambda } + 8 e^{\lambda } \pi p r^2\bigr) + 4 r p'\big]}{4 \rho h r} K_{a}
    -
    \frac{e^{\nu/2}r^{-1-\ell_a}S_{\Omega}(4 + r\nu')}{2\ell_a(\ell_a + 1)a_a \rho h} \nonumber\\
    &\quad
    +
    \frac{e^{\nu/2}r^{-\ell_a}}{2\rho h}\bigg[\frac{ S_t\nu'-\rho h  S_{tr}}{\dot{a}_a} 
    +
    \frac{2\rho h  S_{r\Omega} + \ell_a(\ell_a + 1) S_r + 2 S_{\Omega}'}{\ell_a (\ell_a + 1)a_a }\bigg],
\end{align}
where 
\begin{align}
    S_{tr}&
    \equiv
    \int Y_{\ell m}^{\ast}\bm{S}_{tr}\,d\Omega,
    &S_{r\Omega}
    &
    \equiv
    \int \psi_{\ell m}^{A\ast}\bm{S}_{rA}\,d\Omega,
    \\
    S_{\Omega^2}&
    \equiv
    \int \psi_{\ell m}^{AB\ast }\bm{S}_{AB}\,d\Omega,
    &
    S_{t}&
    \equiv
    \int Y_{\ell m}^{\ast}\bm{S}_{t}\,d\Omega,
    \\
    S_{r}&
    \equiv
    \int Y_{\ell m}^{\ast}\bm{S}_{r}\,d\Omega,
    &
    S_{\Omega}
    &
    \equiv
    \int \psi_{\ell m}^{A\ast}\bm{S}_{A}\,d\Omega.
\end{align}
\end{subequations}
\end{widetext}
The explicit expression of the source through second odrer in perturbation is included in a supplementary \textsc{Mathematica} notebook.

\section{Ratio between Linear and Bilinear Tidal Moments\label{app:second_order_tidal_moments}}
The second-order tidal moment constructed from two quadrupolar tidal tensor is given by
\begin{align}
    d^{(2)}_{2m}=-\frac{2}{3}\mathcal{E}_{\gamma\langle\alpha}\mathcal{E}^{\gamma}{}_{\beta\rangle}\mathscr{Y}^{\langle \alpha\beta\rangle}_{2m}.
\end{align}
Using Eqs.~\eqref{eqn:first_moment} we can rewrite the right-hand side as
\begin{align}
    d_{2m}^{(2)}=-\frac{128\pi^2}{75}\sum_{m'm''}d_{2m'}^{(1)}d_{2m''}^{(1)}\mathscr{Y}_{2m'}^{\ast c\langle a }\mathscr{Y}_{2{m''}}^{\ast{b\rangle}c}\mathscr{Y}_{2m}^{\langle ab\rangle}.
\end{align}
To evaluate the triple product of the STF harmonics, we use~\cite{thorne80}
\begin{align}
    \mathcal{A}_{\langle L_\mathcal{A}\rangle} \mathcal{B}^{\langle L_\mathcal{A} \rangle}=\frac{(2A+1)!!}{4\pi A!}\int \mathcal{A}_{\langle L_\mathcal{A} \rangle} n^{\langle L_\mathcal{A}\rangle }\mathcal{B}_{\langle L_\mathcal{B}\rangle} n^{\langle L_\mathcal{B} \rangle}\,{\rm d}\Omega,
\end{align}
if ${\rm rank}(\mathcal{A})=A={\rm rank}(\mathcal{B})$ for arbitrary STF tensors $\mathcal{A}$ and $\mathcal{B}$. Then triple product evaluates to
\begin{align}
    &\mathscr{Y}_{2m'}^{\ast c\langle a }\mathscr{Y}_{2{m''}}^{\ast b\rangle c}\mathscr{Y}_{2m}^{\langle ab\rangle}\nonumber \\
    &
    =
    \frac{15}{8\pi}\int \mathscr{Y}_{2m'}^{\ast c\langle a }\mathscr{Y}_{2{m''}}^{\ast b\rangle c} n^{\langle a}n^{b\rangle} Y_{\ell m}^\ast \,{ d}\Omega\nonumber \\
    &
    =
    \frac{75}{16\pi} (-1)^{m' + m''}\int \mbf{Y}^{1,2,-m'}\cdot \mbf{Y}^{1,2,-m''}Y_{\ell m}^\ast \,{ d}\Omega,\label{eqn:stf_triple}
\end{align}
where we used
\begin{subequations}
\begin{align}
    \mathscr{Y}^{\langle L\rangle}_{\ell m}n_{\langle L\rangle}&=Y_{\ell m}^\ast,\\
    (-1)^{m}\mathscr{Y}^{\ast \langle L\rangle}_{\ell m}&=\mathscr{Y}^{ \langle L\rangle}_{\ell, -m},
\end{align}
\end{subequations}
and the relation between the STF harmonic and pure orbital harmonic
\begin{equation}
\begin{gathered}
    Y_j^{\ell -1,\ell m}=\sqrt{\frac{\ell}{2\ell + 1}}\mathscr{Y}^{\langle jA_{\ell-1}\rangle }_{\ell m}n_{A_{\ell -1}}, \\
    n_{A_{\ell -1}}=n_{i_1}n_{i_2}\cdots n_{i_{\ell -1}}.
\end{gathered}
\end{equation}
We can evaluate the dot product between the pure orbital harmonics in Eq.~\eqref{eqn:stf_triple} using
\begin{align}
\mbf{Y}^{\ell',\ell,m}=\sum_{m'=-\ell'}^{\ell'}\sum_{m''=-1}^{1}\langle 1\,\ell'm''m'|\ell m\rangle\bm{\mathscr{e}}^{m''}Y_{\ell'm'},\label{eqn:pure_orbital_harm}
\end{align}
where $\langle \ell''\ell'm''m'|\ell m\rangle$ is Clebsch-Gordan coefficient and $\bm{\mathscr{e}}$ is the basis vectors
\begin{align}
    \bm{\mathscr{e}}^{0}\equiv \mbf{e}_z, &&\bm{\mathscr{e}}^{\pm 1}\equiv\mp \frac{1}{\sqrt{2}}(\mbf{e}_x + i\mbf{e}_y).
\end{align}
Plugging in Eq.~\eqref{eqn:pure_orbital_harm} in the last integral of Eq.~\eqref{eqn:stf_triple} yields integrals with triple product of spherical harmonics, which can be computed using Wigner-$3j$ symbols.
Following the procedure outlined above, the second-order tidal moment for $m=0$ reads
\begin{align}
    d_{20}^{(2)}=-\bigg(\frac{16\pi}{5}\bigg)^{1/2}\Big[d_{20}^{(1)2} - 2d_{22}^{(1)}d_{2,-2}^{(1)}\Big],
\end{align}
where we used the fact that the first-order tidal quadrupole is non-zero only for $m\in\{-2,0,2\}$.

\section{Vanishing of Constraint-Fixing Terms at Newtonian Order\label{app:constraint}}
In this section, we show that the contribution of  $\bm{y}_T^{(2)}[\bm{y}_A,\bm{y}_B]$ to the projected equation of motion vanishes at $\mathcal{O}(c^{0})$.
We will keep only the leading terms in $c^{-1}$ and work in flat Cartesian coordinates, where a mode's metric perturbation is
\begin{align}
    h_{\alpha\beta,p}&=- \frac{2\delta\Phi_p}{c^2}\delta_{\alpha\beta}
\end{align}
where $\delta\Phi_p$ satisfies
\begin{align}
    \nabla^2\delta\Phi_p = 4\pi \delta\rho_p= -4\pi \nabla_i(\rho \xi^i_p).
\end{align}
To the same order,
\begin{subequations}
\begin{align}
    \bm{E}_{tt}[\bm{y}_T^{(2)}]&=\frac{2}{c^2}\big(\nabla^2\delta\Phi_T^{(2)}-4\pi  \delta\rho[\xi_T^{(2)}]\big) + \mathcal{O}(c^{-4}),\\
    \bm{E}_{ij}[\bm{y}_T^{(2)}]&= \mathcal{O}(c^{-4}),\\
    \bm{E}_{i}[\bm{y}_T^{(2)}]&= -\bm{C}_i[\xi_T^{(2)}]-\rho\nabla_i\delta\Phi_T^{(2)} + \mathcal{O}(c^{-2}),
\end{align}
\end{subequations}
where $\bm{C}$ is the Newtonian linear pulsation operator (see \emph{e.g.}, Refs.~\cite{chandrasekhar64,lyndenbell1967}).
We exploit the freedom in $\bm{y}_T^{(2)}$ to set $\xi_T^{(2)}=0$, which leads to
\begin{align}
    \bm{C}_i[\xi_T^{(2)}]=0,&&\delta\rho[\xi_T^{(2)}]=-\nabla_i\big[\rho \xi_T^{(2)i}\big]=0.
\end{align}

Also, the decomposition in Eq.~\eqref{eq:second-order-induced-equations} fixes $\nabla^2\delta\Phi_T^{(2)}$ to represent the second-order residual in the Poisson's equation.
Let us start from the total gravitational potential that satisfies
\begin{align}
    \nabla^2\delta\Phi^{(2)}&=4\pi \delta\rho^{(2)},
\end{align}
where, from Sec.~\ref{sec:formalism},
\begin{align}
    \delta\rho^{(2)}&=-\nabla_i\big[\rho \xi^{(2)i}\big] + \nabla_i\big[\xi^{(1)i}\nabla_j(\rho\xi^{(1)j})\big],\nonumber \\
    &\equiv -\nabla_i\big[\rho \xi^{(2)i}\big] + \tilde{\rho}.\label{eqn:density_perturbation_constraint}
\end{align}
Equation~\eqref{eq:second-order-induced-equations} maps the first term in Eq.~\eqref{eqn:density_perturbation_constraint} onto the mode basis. 
Therefore, we have
\begin{align}
    \nabla^2\delta\Phi_T^{(2)}
    =
    4\pi \tilde{\rho}.\label{eqn:second_order_tidal_field_poisson}
\end{align}
Since $\tilde{\rho}=0$ outside the star, 
\begin{align}
    \delta\Phi_T^{(2)}(r> R)\sim r^{-({\ell + 1})}.
\end{align}

Now, we evaluate the contribution of the second-order tidal field in the equation of motion,
\begin{align}
    &\int \bigg(\xi_{\alpha,p}\bm{E}^\alpha[\bm{y}_T^{(2)}] + \frac{h_{\alpha\beta,p} c^4}{16\pi }\bm{E}^{\alpha\beta}[\bm{y}_T^{(2)}]\bigg)\,d^3x
    \nonumber \\
    &=-\int\bigg[ \rho\xi_{\alpha,p}\nabla^\alpha \delta\Phi_T^{(2)} +  \frac{1}{4\pi } \delta\Phi_p \big(\nabla^2 \delta\Phi^{(2)}_T\big)\bigg]\,d^3x.\label{eqn:second_order_tidal_field_eom}
\end{align}
Substituting Eq.~\eqref{eqn:second_order_tidal_field_poisson} in Eq.~\eqref{eqn:second_order_tidal_field_eom} and integrating by parts using $\rho(r\geq R)=0$, we have
\begin{align}
    &\int \bigg[\delta\rho_p \delta\Phi_T^{(2)}-\frac{1}{4\pi }\delta\Phi_p\nabla^2 \delta\Phi_T^{(2)}\bigg]d^3x
    \nonumber \\
    &=\iint\bigg[
    -
    \frac{\delta\rho_p(\bm{x})\tilde\rho(\bm{y})}{|\bm{x}-\bm{y}|}
    +
    \frac{\delta\rho_p(\bm{y})\tilde\rho(\bm{x})}{|\bm{y}-\bm{x}|}\bigg]\,d^3xd^3y
    \nonumber \\
    &=0.
\end{align}
Hence, at order $\mathcal{O}(c^{0})$, $\bm{y}_T^{(2)}[\bm{y}_A,\bm{y}_B]$ has a vanishing contribution to a mode's equation of motion.